# Hypersonic Weapons-Part 1:
# Background and Vulnerability to Missile Defenses

David Wright[1] and Cameron L. Tracy[2]

February 2, 2023

Assessing the desirability of hypersonic glide vehicles (HGVs) requires understanding what capabilities they may provide compared to alternative systems that could carry out the same missions, particularly given the technical difficulties and additional costs of developing hypersonic gliders compared to more established technologies. Part 1 of this paper considers the primary motivations given for developing HGVs and summarizes the current development programs of the United States, Russia, and China. It then analyzes the vulnerability of hypersonic boost-glide vehicles (BGVs) to missile defenses and finds that evading the most capable current endo-atmospheric defenses requires that BGVs maintain speeds significantly higher than Mach 5 throughout their glide phase. Achieving these high speeds has important implications for the mass and range of these weapons, which in turn affect what delivery vehicles can carry them and how their performance compares to other means of delivery, as is discussed in Part 2.

The United States, Russia, and China are leading the development of hypersonic weapons: missiles primarily intended to glide for long distances within the atmosphere at speeds greater than five times the speed of sound, or Mach 5.[3] These include hypersonic cruise missiles, which carry engines to power them during flight, and boost-glide vehicles, which are not powered during their glide phases. Several other nations have active, smaller-scale development programs.[4]

Assessing the desirability of developing hypersonic glide vehicles requires understanding what capabilities they may provide compared to alternative systems that could carry out the same missions, particularly given the technical difficulties and additional costs of developing hypersonic gliders compared to more established technologies.

In particular, it is useful to compare hypersonic glide weapons to ballistic missiles flown on depressed trajectories (DTs) and carrying maneuverable reentry vehicles (MaRVs).[5] DTs refer to ballistic trajectories with significantly smaller loft angles, and therefore lower apogees and shorter pathlengths, than those that maximize the range of a ballistic missile. A MaRV, which uses atmospheric forces to maneuver during the terminal phase of missile flight, can provide many of the key capabilities motivating hypersonic weapon development—in particular, the ability to maneuver during reentry could allow it to dodge terminal defenses, to use terminal guidance for high accuracy, to retarget over hundreds of kilometers, and to dive to its target at a steep angle. Unlike hypersonic gliders, MaRVs cannot maneuver significantly during midcourse flight or fly for long distances at altitudes below 50 km, and do not take advantage of glide to substantially increase their flight ranges.

In our previous paper, referred to here as Paper 1,[6] we presented an analysis of hypersonic weapon performance using a computational model of hypersonic flight and vehicle parameters derived from the US tests of the Hypersonic Technology Vehicle-2 (HTV-2), a representative glider, conducted in 2010 and 2011.[7] We assessed common assertions made about hypersonic weapons with regard to delivery time, maneuverability, and detectability, showing that many of these claims were exaggerated or false.

Paper 1 focused primarily on long-range hypersonic vehicles, like the US HTV-2 and the Russian Avangard (Vanguard), which is now deployed on SS-19 long-range missiles carrying nuclear warheads (although its capabilities are not known).[8] Both systems are said to reach speeds of Mach 20 or higher.[9]

Since testing the HTV-2 in the early 2010s, US focus shifted from intercontinental missiles to shorter-range, lower-speed systems. While the analysis in Paper 1 also applied to shorter range hypersonic weapons, in this paper we consider issues relevant to these shorter-range systems more specifically.

Since MaRVs launched on ballistic missiles with ranges of more than a few hundred kilometers travel at hypersonic speeds, the term "hypersonic weapons" is ambiguous. We instead refer to weapons that travel a significant portion of their trajectories at low altitudes, relying primarily on lift forces to stay aloft, as "hypersonic glide vehicles" (HGVs). HGVs include both boost-glide vehicles (BGVs), which are accelerated to high speed by rocket boosters and then glide without power, and hypersonic cruise missiles (HCMs), which also use boosters to reach hypersonic speed but carry engines to power them during flight.

In this paper, we first discuss the primary motivations given for developing hypersonic weapons and then summarize the evolution and current status of US BGV programs. We then look in detail at three topics:

(1) The capability of current missile defense systems to intercept hypersonic weapons. An important implication of this analysis is that evading defenses appears to require that BGVs maintain speeds throughout their glide phases that are significantly faster than the speed necessary to be considered "hypersonic" (i.e., having speeds greater than Mach 5).

(2) A comparison of the ranges, masses, and flight times of BGVs and MaRVs under various conditions, including when the BGV speed is high enough to evade defenses.

(3) An analysis of the reductions of speed and range caused by midcourse maneuvering by BGVs. Such maneuvering is cited as an advantage of BGVs over MaRVs, so understanding the tradeoffs of such maneuvers is important for comparing the two systems.

We also briefly discuss the issue of whether BGVs or MaRVs will develop a plasma sheath during reentry. This issue is important since plasma can interfere with communication to the vehicle that might be important for achieving high accuracy.

**Motivations for Hypersonic Weapons**

Advocates of hypersonic weapons have discussed a broad set of motivations for developing and deploying these vehicles; we list key motivations below.[10] There has been little discussion, however, of which of these motivations can uniquely, or even best, be fulfilled by weapons traveling at hypersonic speeds at low altitudes (i.e., HGVs) compared to similarly fast ballistic missiles or slower supersonic cruise missiles.

It therefore remains unclear whether these motivations justify the efforts and expenditure needed to solve the difficult technical challenges facing HGV development.[11] In other words, do HGVs offer meaningful advantages over systems that use more well-developed technologies, such as ballistic missiles with MaRVs?

The primary motivations for US development of hypersonic weapons given by those leading the development are:[12]

1. To quickly attack targets from long ranges (which requires short delivery times)
2. To achieve high accuracies and home on targets (which requires terminal maneuvering)
3. To retarget over a large area during flight (which requires terminal and/or midcourse maneuvering)
4. To evade or destroy air defenses and terminal missile defenses.[13]

If the aim is to accomplish goals (1) – (4), there are several different trajectories a vehicle could fly to do so, using different types of weapons.

For example, these goals do not require flying low in the atmosphere for long distances and could instead be carried out by ballistic missiles with MaRVs. MaRVs have the significant advantage that they rely largely on existing technology, and therefore present fewer technical challenges and may therefore have higher reliability than BGVs and HCMs. In addition, MaRVs launched on depressed ballistic trajectories can have shorter delivery times than HGVs because they encounter lower drag forces.[14] This means that the desire for hypersonic speed and short delivery time does not argue uniquely for HGVs.

Similarly, if one prioritizes maneuvering over traveling at hypersonic speed, then developing systems like supersonic cruise missiles would be easier. For example, a supersonic vehicle traveling at Mach 3 would be subjected to a heating rate during its glide phase that is roughly eight times smaller than a hypersonic vehicle traveling at Mach 6, since atmospheric heating increases approximately with the cube of velocity.

This paper compares the capabilities of BGVs and MaRVs designed to reach distances of up to a few thousand kilometers and speeds of Mach 5 to 13. While HCMs are not considered in detail, the models developed here apply to analysis of these systems as well. We show that making such a comparison requires specifying the mission the weapons are intended to carry out, the launch platform, the need to evade defenses, and other parameters. For many proposed missions, ballistic missiles equipped with MaRVs would outperform BGVs launched from the same platform.

For our modeling below, it is important to recognize that the dynamics of a BGV during glide (velocity profile, range, and flight time) depend almost entirely on the lift-to-drag ratio $L/D$ and very weakly on the ballistic coefficient $\beta$ for a given initial glide speed $V$ and vehicle mass $m$. Appendix A shows that changing $\beta$ will change the drag at a given altitude but will also change the glide altitude in a way that the vehicle feels essentially the same drag force. This implies that Acton's fitting to the test data of the HTV-2 constrained the value of $L/D$ much more tightly than the value of $\beta$.[15] But since the dynamics depend only weakly on $\beta$, uncertainties in the value of $\beta$ have little effect on the analysis discussed in this paper and in Paper 1.[16]

## Comparing HGVs with MaRVs

BGVs and MaRVs are not entirely distinct means of delivering warheads but lie on a spectrum; both operate in similar ways, differing mainly in the extent to which they rely on lift versus inertial forces to fly long distances. A standard ballistic missile reentry vehicle relies on high speed to reach distant targets, while gliders sacrifice some speed to drag and instead rely on lift forces by travelling at lower altitudes for a significant portion of flight. MaRVs represent a middle ground, flying much of their trajectory ballistically, then using lift forces to maneuver late in flight. A MaRV might be considered a hypersonic glider that reenters the atmosphere relatively late in flight; a hypersonic glider might be considered a MaRV that reenters earlier. Both MaRVs and HGVs follow "non-ballistic" trajectories for a portion of their flight.

For long-range systems, especially those with intercontinental ranges, the distinction between a BGV and a ballistic missile with a MaRV is clearer. Both are launched on rocket boosters, but the BGV will spend a much larger fraction of its trajectory within the atmosphere. For example, the full test range for the US HTV-2 flight tests was about 7,600 km. The glider was intended to travel about 75 percent of that distance at altitudes below 100 km (generally considered to be the upper edge of the atmosphere) and about 60 percent gliding at altitudes below 50 km.[17] In contrast, a MaRV reentering from a long-range ballistic trajectory would spend well under 10 percent of its trajectory below 50 km altitude.

For shorter ranges, however, the distinction is less clear. The apogee of an efficient ballistic missile trajectory decreases as its maximum range decreases, so the reentry vehicle will spend a larger fraction of its trajectory within the atmosphere for a shorter trajectory. A warhead on a 10,000 km-range "minimum-energy" trajectory (MET) will reach altitudes of about 1,500 km.[18] In contrast, missiles on METs with ranges of less than 400 km never leave the atmosphere (for short ranges the apogee of a ballistic missile on an MET is about a quarter of its range). In addition, ballistic missiles flown on less energy-efficient depressed trajectories can use a more powerful booster to reach the same range as a warhead following an MET, but with a lower apogee due to their higher initial speed.[19]

A potential advantage of gliding is that by using lift forces to stay aloft a BGV can reach a given range with a lower initial speed than a ballistic missile warhead and could therefore require a smaller booster to accelerate it. Booster technology, however, is very well developed and building a somewhat larger booster to deliver a ballistic missile warhead is often feasible. Using a smaller booster could be an advantage for vehicles launched from aircraft since it would reduce the amount of mass the aircraft would need to carry; they could also be useful for vehicles launched from submarines where space is restricted. We analyze the issue of booster mass in Part 2 of this paper.

The fuzziness in the distinction between MaRVs and BGVs is apparent in the history of US development efforts, discussed in the next section.[20]

**Evolution of the US Hypersonic Program**

After roughly half a century of intermittent work on hypersonic vehicles, US interest in developing HGVs grew following the September 11, 2001, terrorist attacks.[21] They were seen as a way to deliver prompt, global attacks with conventional (non-nuclear) weapons against threats such as terrorist organizations and activities.[22] While ballistic missiles could be modified for this mission, long-range ballistic missiles in the US arsenal are designated solely for nuclear warhead delivery. Using HGVs was seen as a way of reducing the chance that the launch of a long-range strike might be construed as the beginning of a nuclear attack, prompting a nuclear exchange. This interest led to the development and testing of the long-range HTV-2 vehicle as part of the Force Application and Launch from CONtinental United States (FALCON) program, announced in 2003 and run by the Defense Advanced Research Projects Agency (DARPA).

The HTV-2 glide vehicle was based in part on "waverider" designs first proposed in the early 1950s.[23] These designs use a wedge or half-cone shape intended to match the shock-wave pattern of the airflow around the glider, enclosing part of the shock wave under the vehicle to provide additional lift. Since the shock pattern depends on the specific speed and altitude of the vehicle, this design concept could not be applied directly to long-range boost-glide vehicles, since their speed and altitude can change significantly during the glide phase.[24] Drawing on this design, however, was seen as a way of improving the performance of the vehicle by increasing its lift-to-drag ratio (*L*/*D*). Plans called for the HTV-2 to have *L*/*D* of 3.5 to 4 and for the follow-on HTV-3 to have *L*/*D* of 4 to 5. The HTV-2 in fact demonstrated *L*/*D* of about 2.6 in tests, and the program was terminated before an HTV-3 vehicle was produced (see Appendix A: Implications of Increasing *L*/*D*).[25]

Around 2003 the Department of Defense (DoD) started a second, parallel track for developing a long-range hypersonic weapon, called the Advanced Hypersonic Weapon (AHW).[26] It was based on an older MaRV design called the Sandia Winged Energetic Reentry Vehicle Experiment (SWERVE) that originated in the late 1970s and was flight tested through the mid-1980s.[27] Like the SWERVE, the AHW had a conical body with fins, rather than the wedge shape of waverider-like vehicles. DoD officials saw this design as a more technologically mature backup to the FALCON program.[28] The AHW had a single successful test, in 2011.[29]

By the mid-2010s, US focus shifted to shorter-range hypersonic weapons with speeds of Mach 5 to 10 and ranges of a few thousand kilometers. This shift was driven in part by ongoing difficulties with the HTV-2 tests, some of which were related to the severe heating from long-distance flight at speeds near Mach 20.[30] In addition, the US mission for hypersonic weapons changed: Use in theater-scale conflicts to evade or destroy defensive systems became a key motivation.

After ending the HTV program in the mid-2010s, DARPA began to develop a short-range BGV with a wedge design called the Tactical Boost Glide (TBG) system (Figure 1). Around the same time the United States began developing a new short-range BGV with a conical design, based on the AHW. In 2018, the DoD announced that its goal was to produce a conical vehicle, called the Common Hyper Glide Body (C-HGB), for joint use by the Army, Navy, and Air Force, with each service launching it on its own delivery system (Figure 2). The Army program is the Long-Range Hypersonic Weapon (LRHW) and the Navy program is the Conventional Prompt Strike (CPS) weapon.[31] The Air Force version was the Hypersonic Conventional Strike Weapon (HCSW), but the Air Force ended this program in 2020.[32]

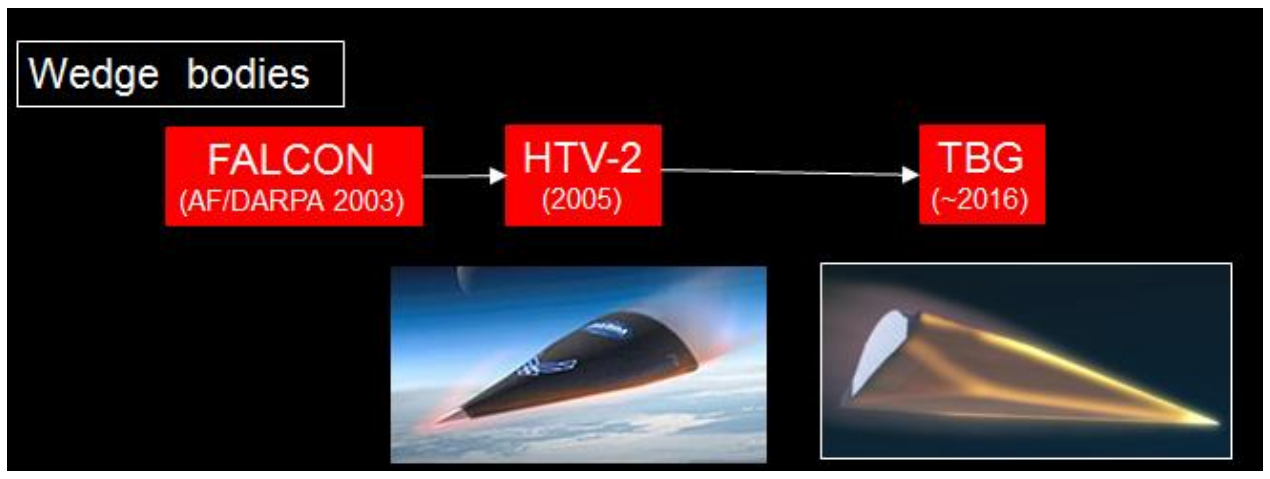

Figure 1. The evolution of BGVs using wedge designs that draw on waverider development.

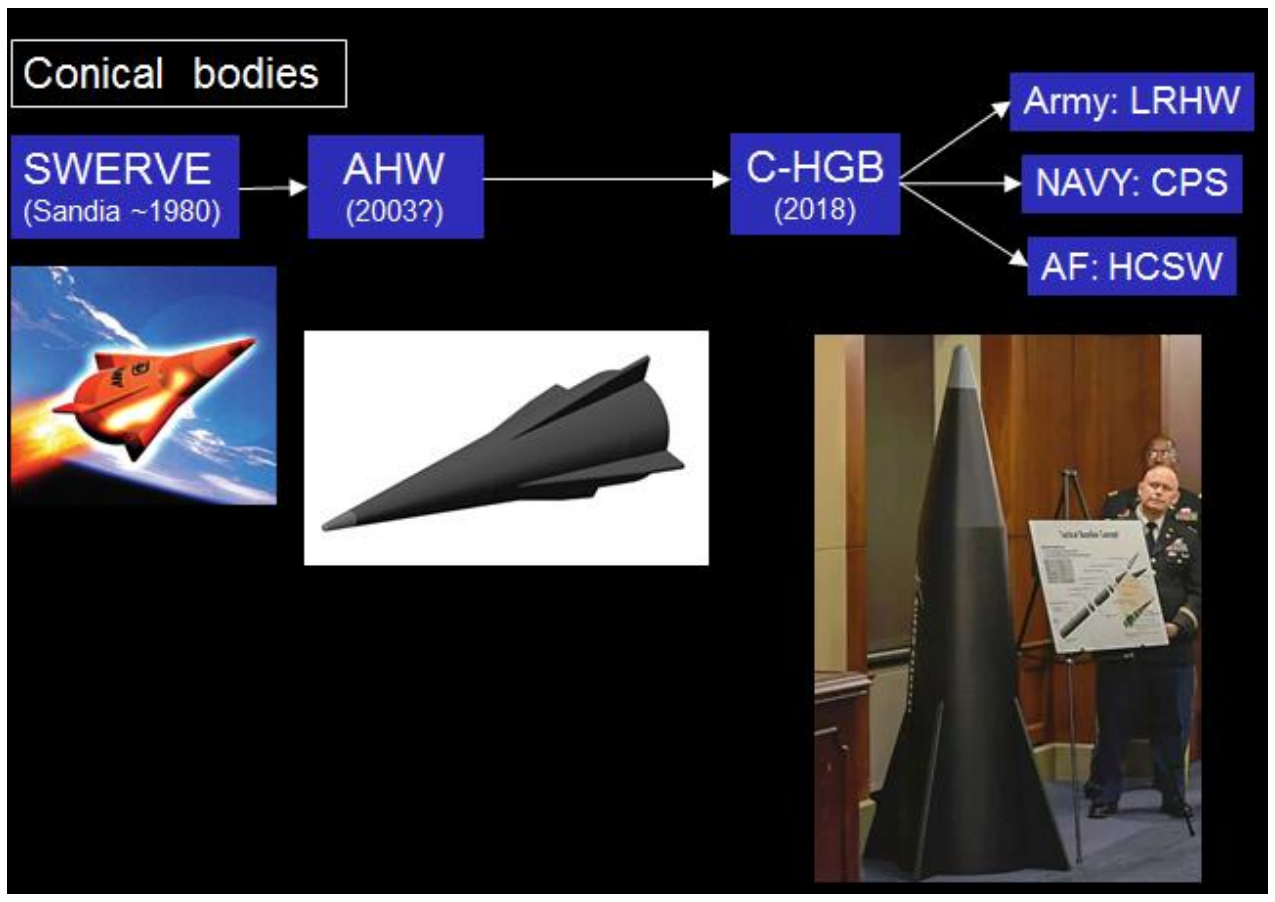

Figure 2. The evolution of BGVs using conical designs that draw on MaRV development.

The C-HGB, as a direct descendent of the SWERVE and AHW, is clearly derived from MaRV designs. It has a conical design and an estimated *L*/*D* of about 2.2—lower than that of the HTV-2, which had *L*/*D* of 2.6.[33] As a result, the DoD's decision to pursue a more mature technology will likely result in a BGV with less range and maneuverability than a successful wedge design with higher *L*/*D* would have.[34] This also means that the computational model we developed in Paper 1, based on the wedge design of the HTV-2, may somewhat over-estimate the flight performance of the C-HGB.

In 2019, the Air Force announced that it was leaving the C-HGB program to work with DARPA on the TBG program to develop an air-launched BGV called the Air-launched Rapid Response Weapon (ARRW). The Air Force reportedly saw the TBG-derived ARRW as a potentially lighter vehicle than the C-HGB, which would allow an aircraft to carry more of them (Figure 3).[35]

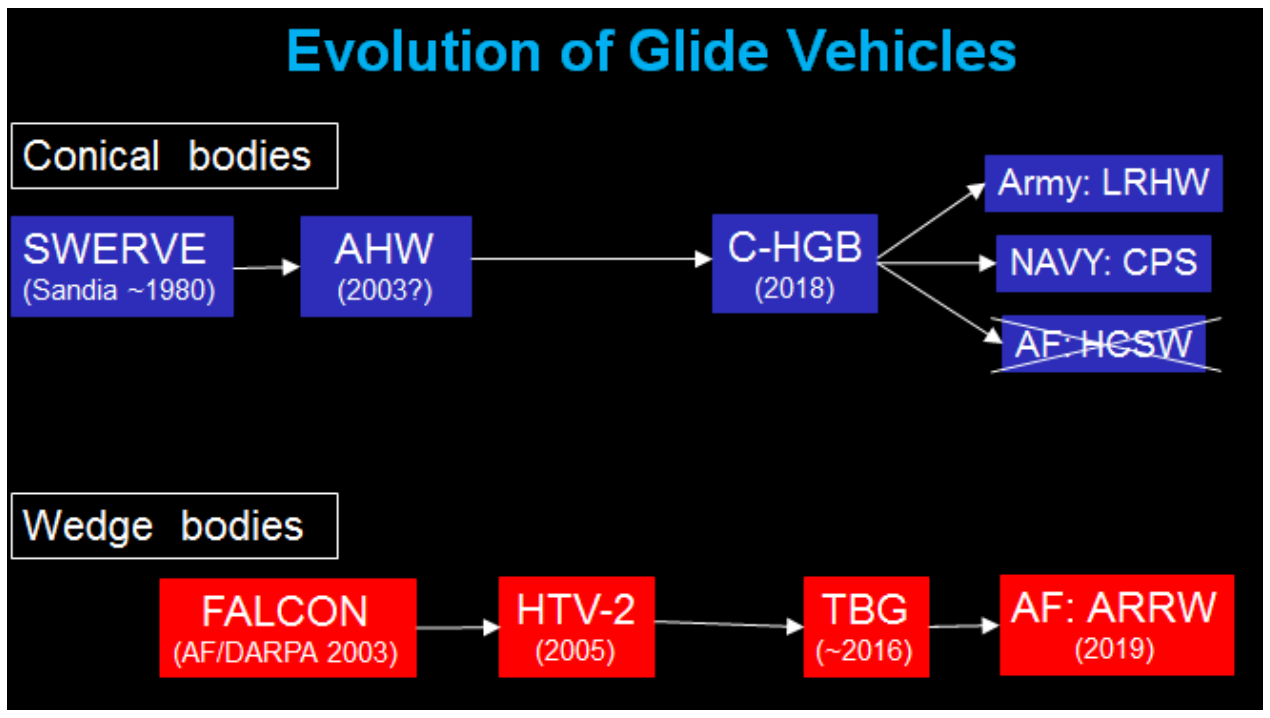

Figure 3. The evolution of current US BGV programs.

At the same time as this development of BGVs, the United States and other countries have been developing HCMs powered by air-breathing engines called scramjets, with speeds in the range of Mach 5 to 10 and ranges of up to a couple thousand kilometers. From 2009 through 2013, the United States tested the X-51A HCM, which reached Mach 5.1. An ongoing joint DARPA/Air Force program to develop the Hypersonic Air-breathing Weapon Concept (HAWC) HCM builds on the X-51 program.[36]

Military HCM programs use hydrocarbon fuels for their scramjets, like the X-51, rather than hydrogen fuel, like the US NASA X-43 vehicle.[37] The use of this fuel limits their speed to below about Mach 8 to 10.[38] In 2020 Mark Lewis, then director of modernization at the DoD's Research and Engineering office, said he expects that air-breathing systems will "probably top out around Mach 7."[39] This relatively low speed may limit the utility of HCMs for certain missions, as discussed below.

In recent decades, US hypersonic development has proceeded in parallel with large Russian and Chinese development programs. In addition to Russia's long-range Avangard BGV, it has also tested a shorter-range ship-based glider, called the Tsirkon (Zircon), which is an HCM carrying a scramjet. It is reported to reach speeds of Mach 6 to 8 and ranges of 400 to 1,000 km.[40] As noted above, the Russian Kinzhal (Dagger) weapon has been described as an air-launched ballistic missile with fins that maneuvers during reentry, rather than a BGV. It is reported to reach speeds of Mach 10, with a range up to about 1,500 km.[41]

China is reportedly working on two HGVs. The DF-ZF (previously called WU-14) is a BGV carried on a DF-17 booster, and reportedly able to reach speeds of Mach 5 to 10 and a range of 1,200 km.[42] China declared it operational in October 2019.[43] In addition, China is developing an HCM called the Xingkong-2 (Starry Sky II), which uses a waverider design and carries a scramjet. During its first test in August 2018, it was reported to have reached speeds of Mach 5.5 to 6, traveling at 29 to 30 km altitude for 400 seconds, which gives a glide range of 660-720 km.[44] Some reports suggest the vehicle could be operational in the mid-2020s.[45]

As discussed above, a key motivation for developing HGVs, particularly the relatively short-range systems the United States is currently focusing on, is to evade terminal missile defenses. Whether an HGV can do so depends on its speed and design, as discussed in the next section. Understanding that issue is important for comparing BGVs and MaRVs.

**The Capability of Missile Defenses against Hypersonic Weapons**

This section assesses the ability of existing missile defense systems to intercept hypersonic weapons and identifies the characteristics a weapon needs to evade interceptors.

The first parameter that governs the vulnerability of a vehicle to a specific defensive system is its flight altitude relative to the operational altitude window of the interceptor. When HGVs are in the glide portion of their flight, their equilibrium glide altitude is determined by their speed, mass, and lift coefficient, or alternately by their speed, $L/D$, and ballistic coefficient $\beta$ (see Equation 19 below). Given current values of $L/D$ of 2 to 3, gliders will typically fly at about 26 km altitude at Mach 5 (1.5 km/s) and at about 43 km at Mach 15 (4.5 km/s). If $L/D$ were increased to 6, these altitudes would increase to about 32 km at Mach 5 and 50 km at Mach 15.[46]

Once wedge-shaped HGVs approach their target, they can leave the glide phase by flipping upside down and using their lift forces to dive sharply toward the Earth.[47]

Two types of missile defenses are currently deployed. Midcourse defenses, such as the US Ground-based Midcourse Defense (GMD) and ship-based Aegis Standard Missile-3 (SM-3) systems, are designed to engage weapons at long distances and high altitudes when they are traveling on predictable trajectories above the atmosphere. Such "exo-atmospheric" defenses are restricted to operate at altitudes of 100 km or higher and in principle can defend large ground areas.

Terminal defenses, on the other hand, attempt to engage weapons late in flight when they are reentering the atmosphere above their target. These defenses, including the US Patriot and Aegis SM-6, and Russia's S-400 and recently deployed S-500, maneuver aerodynamically and are restricted to operating at tens of km altitude, meaning they are "endo-atmospheric."[48] They engage weapons at short ranges and protect at most small ground areas—perhaps tens of km across. (The US THAAD system is designed to operate in the high-endo to low-exo-atmospheric regions—see below.)

More information is available about US defensive systems than those of other countries. For this analysis we therefore use parameters relevant to US systems and assume those are representative of what other countries deploy now or might deploy in the near future.

*GMD and SM-3 interceptors*

The United States deploys two long-range interceptors intended to protect large areas of the Earth's surface from ballistic missiles: the Ground-Based Interceptors (GBI) deployed in Alaska and California as part of the GMD system, and the SM-3 interceptors deployed on Navy ships as part of the Aegis Sea-Based Midcourse system. Both are hit-to-kill systems, meaning that the interceptor, or kill vehicle, they carry attempts to destroy a target by physically colliding with it. Both interceptors are designed to intercept above the atmosphere and cannot engage targets below about 100 km altitude.[49] Below those altitudes the kill vehicle's infrared (IR) sensors used for homing on the target would be blinded by heating resulting from the kill vehicle's high speed through the atmosphere. As a result, neither type of interceptor could engage HGVs once those vehicles are in their glide phases.

*THAAD*

The Terminal High Altitude Area Defense (THAAD), currently deployed in several nations, is designed to engage ballistic missile reentry vehicles late in the warhead's flight as it is beginning to reenter the atmosphere, between about 40 and about 150 km altitude. The THAAD interceptor has a speed of about Mach 9 (2.7 km/s) and uses an IR sensor with a sapphire window to limit the blinding effects of atmospheric heating; this restricts its use to altitudes above 40 km.[50] The kill vehicle is a hit-to-kill system. It has two sets of thrusters for maneuvering: a set of four near the center of the kill vehicle to provide divert forces at high altitudes, and a set of six near the back of the vehicle for attitude control, which can also be used for maneuvering at lower altitudes by creating atmospheric lift forces.[51]

Since HGVs flying at speeds less than about Mach 10 would have equilibrium glide altitudes below about 40 km, THAAD could not engage them during glide.

THAAD might be able to engage either HGVs or MaRVs as they were reentering the upper atmosphere at altitudes where the atmospheric density was low enough that these weapons could not generate sufficient lift forces to out-maneuver the interceptor. Since its operational range is only about 200 km, however, THAAD interceptors co-located with the weapon's target would not be able to reach an HGV before the start of its glide phase and might not be able to reach a MaRV on a depressed trajectory before its altitude was too low for THAAD to engage.

Moreover, at high altitudes THAAD would be vulnerable to decoys and other countermeasures similar to those designed to penetrate purely exo-atmospheric defenses.[52] These countermeasures would not work at lower altitudes but could prevent THAAD from identifying the warhead and launching early enough to intercept above 40 km.

As a result, it appears that THAAD would not be able to reliably intercept hypersonic weapons, leaving only defenses with purely endo-atmospheric interceptors to try to engage them.

*Patriot and other Endo-atmospheric Interceptors*

Current endo-atmospheric interceptors use aerodynamic forces to maneuver to hit missile warheads late in their trajectories when they are low in the atmosphere—below about 40 km altitude. Because they intercept so late, each interceptor can defend a much smaller ground area than the longer-range interceptors discussed above—typically a region a few tens of kilometers in radius. They can therefore attempt to defend small targets like some military installations, but covering a large, populated region would require many interceptors, which would have to be interspersed within the region.

Current US endo-atmospheric systems include versions of the Patriot Advanced Capability-3 (PAC-3) defense, as well as SM-2 and SM-6 interceptors deployed on Navy ships.

During typical exo-atmospheric engagements, the only force acting on the object the interceptor is trying to hit is gravity, and the dynamics of the intercept depend only on the relative speed of the interceptor and target—the closing speed—rather than the individual speeds of each object. For endo-atmospheric engagements, however, both the interceptor and target can use the atmosphere to maneuver, and what matters is the relative lateral acceleration that the two objects can achieve at the altitude of the engagement as each attempts to outmaneuver the other. In other words, the interceptor must be able to closely match any evasive movement of the target.

An important principle of guidance and control theory is that real-world interceptors must be able to achieve two to three times the lateral acceleration of a maneuvering target to reliably intercept it.[53] The estimated miss distance for an engagement between an interceptor and target will depend on specifics of the interceptor sensor and guidance system, as well as what the defense knows about the target and the kind of maneuvers the target may execute. The target will have a good idea of what maneuvers it can do that will be most stressing for the interceptor.[54]

The lateral acceleration $a$ of an object with mass $m$ and lift coefficient $C_L$ and traveling with velocity $V$ at an altitude with atmospheric density $\rho$ is given by:

$$a = \frac{C_L A \rho V^2}{2m} \quad (1)$$

where $A$ is a characteristic area associated with the body.

The relative acceleration of a target and interceptor at the same location, and therefore the same $\rho$, is:

$$\frac{a_{target}}{a_{Int}} = \left(\frac{m_{Int}}{m_{target}}\right)\left(\frac{(C_L A)_{target}}{(C_L A)_{Int}}\right)\left(\frac{V_{target}}{V_{Int}}\right)^2 \quad (2)$$

Note that the most important variable is the velocity ratio of the two vehicles at intercept, since it enters as a square.

Both the target and interceptor will generate lift forces by creating an angle-of-attack between their body axis and velocity vector. At higher altitudes where the atmospheric density and therefore the lift force is relatively low, the maximum lateral acceleration will be set by constraints on how large an angle-of-attack can be achieved, while at lower altitudes where the lift force is high the maximum lateral acceleration will be limited by the ability of the body to withstand the associated forces;[55] for this analysis of relative maneuverability we assume the interceptor and target have similar structural constraints and do not consider them explicitly.

*Lateral Acceleration of an Interceptor and a BGV*

As a relevant example of applying Equation 2, we compare the maneuvering capability of a representative endo-atmospheric interceptor to that of a BGV design. We estimate the mass, lift coefficient, area, and maximum velocity of the PAC-3 Missile Segment Enhancement (MSE) interceptor, which was fielded in 2016 as an advanced version of the PAC-3 interceptor and is currently one of the most capable endo-atmospheric interceptors (see below). With the development of the MSE variant, the original PAC-3 is now called PAC-3 Cost Reduction Initiative (CRI).

For a hypersonic vehicle, we consider a system similar to the HTV-2 and about which aerodynamic parameters are known: a model of the Common Aero Vehicle (CAV), for which lift and drag coefficients have been calculated in a report by Phillips.[56] While a useful model for analyzing hypersonic vehicles, the CAV may not be representative of conical designs like the C-HGB, which might have lower *L*/*D* values and thus less maneuverability.

*PAC-3 MSE parameters*

The PAC-3 MSE has a two-pulse motor. The second pulse might be used as a sustainer to maintain the interceptor's speed during flight (for this analysis we assume the interceptor travels at its maximum speed throughout its flight). The body diameter is 29 cm (compared to 25.5 cm for CRI) and its length is 5.3 m (compared to 5.2 m for CRI), giving a length-to-diameter ratio of about 18.[57] The cross-sectional area of the body is 0.066 $m^2$, and the planform body area (without fins) is about 1.5 $m^2$ (Figure 4).

The MSE is a hit-to-kill interceptor and has a "lethality enhancer," which is a set of rods that shoot out from the body to effectively give the weapon a larger lethal diameter.[58] While this increases somewhat the miss distance that will still allow the interceptor to hit its target, that increase is not large enough to change the dynamic analysis below.

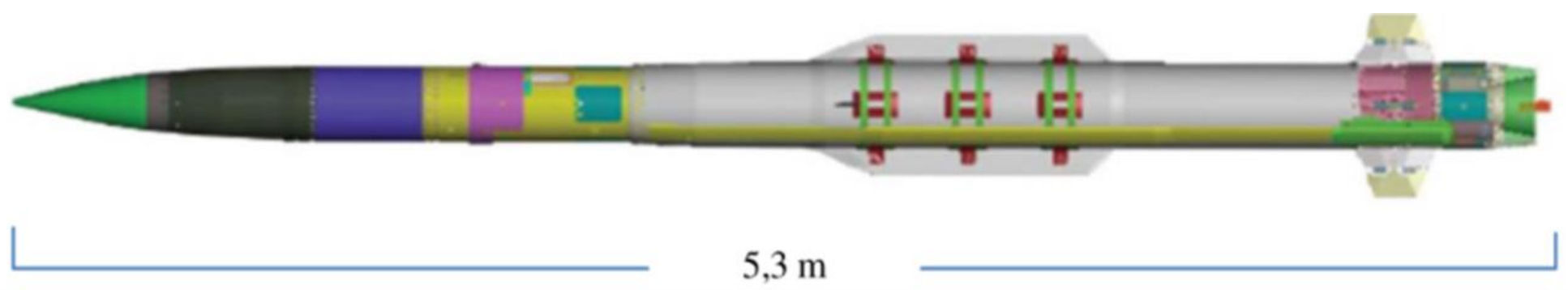

Figure 4: A schematic of the PAC-3 MSE interceptor. (Source: https://web.archive.org/web/20071019124830/http:/www.lockheedmartin.comα/products/PAC3MissileSegmentEnhancement/index.html )

The CRI is reported to have a mass of about 315 kg with 158 kg of propellant, which implies a mass after burnout of ~160 kg; this is the mass we assume for the MSE.[59]

Based on the reported increase in intercept range of the MSE over the CRI, we estimate the MSE speed is about 1.8 km/s (Mach 6), compared to 1.4 km/s (Mach 4.7) for CRI.[60] The MSE is reportedly able to intercept at altitudes of 30+ km, compared to 20+ km for CRI.[61]

Since it is cylindrical, at zero angle-of-attack (φ) the MSE body will generate zero lift. To maneuver it uses a set of small thrusters around the body to give it a non-zero angle-of-attack, which creates lift. To compare the relative values of lift coefficients for the various systems considered here, we estimate the lift coefficients for φ in the range of 10 to 20 degrees for each system.

To estimate the lift coefficient of MSE, we estimate the normal force coefficient $C_N$ of a body with cylindrical symmetry using the equation: [62]

$$C_N = \frac{A_b}{A} \sin 2\varphi \cos\frac{\varphi}{2} + \eta\, c_{d_n} \frac{A_p}{A}\, sin^2\varphi \quad (3)$$

where $A_b$ is the body base area, $A_p$ is the planform area, and $A$ is referred to as a reference area of the body. For high Mach number, the parameters $c_{dn} \sim 1.3$ and $\eta \sim 1$, and are related to the crossflow drag of a cylinder.[63] For small angles-of-attack the lift coefficient $C_L$ is essentially equal to $C_N$.[64]

The quantity that enters the acceleration equation (Equation 1) is $C_LA$, which is given in terms of the two known areas $A_b$ and $A_p$. Using $A_p = ld$ and $A_b = \pi d^2/4$, where $l$ is the body length and $d$ its diameter, Equation 3 can be rewritten for high Mach number as:

$$C_LA \approx C_NA = A_b\left[\sin 2\varphi \cos\frac{\varphi}{2} + \frac{5.2}{\pi}\frac{l}{d}\, sin^2\varphi\right] \qquad (4)$$

Applying Equation 4 to the MSE with $l/d = 18$ gives:

Table 1. Estimated lift values for the MSE interceptor at varying angle-of-attack.

| φ | 10º | 15º | 20º |
|---|---|---|---|
| $C_L A$ | 0.079 | 0.17 | 0.27 |

*BGV parameters*

Data needed to determine the variation of the HTV-2's lift coefficient with its angle-of-attack is not publicly available. To estimate this variation, we compare this vehicle with the CAV, which has a similar design. For the HTV-2 we assume $L/D$ = 2.6 and $m$ = 900 kg.[65]

In his analysis of hypersonic vehicles, Phillips presents aerodynamic parameters for two CAV models with different values of $L/D$, called the CAV-H and CAV-L, for "high" and "low" performance respectively.[66] The CAV-H model in Phillips' report has a mass of 910 kg, a length of 2.7 to 3.7 m, and a diameter of 1.2 m, so it is similar in size to the HTV-2.[67] While it has a somewhat higher value of $L/D$ than the HTV-2 (3.0 to 3.3 at Mach 20 for φ in the range of 10 to 15 degrees) we use its variation in lift values as a function of φ to estimate the approximate performance of related wedge-shaped gliders, including the HTV-2.[68] The $L/D$ of the CAV-H model is maximized near φ = 10° for speeds above about Mach 6.

To compare the maneuverability of the glider to that of the interceptor, we estimate its lift at Mach 6 using the values Phillips calculates for the CAV-H, which gives:

Table 2. Estimated lift values at Mach 6 for Phillips' model of the CAV-H using a reference area $A$ = 750 in$^2$ = 0.48 m$^2$.

| φ | 10º | 15º | 20º |
|---|---|---|---|
| $C_L A$ | 0.20 | 0.33 | 0.47 |

*Implications for Intercept Probability*

Using Equation 2 and the requirement that the interceptor must generate two to three times the lateral acceleration of the target to reliably intercept it, we find the following requirement for the interceptor velocity $V_{int}$:

$$V_{int} \geq \left[(2\ to\ 3)\left(\frac{m_{Int}}{m_{target}}\right)\left(\frac{(C_L A)_{target}}{(C_L A)_{Int}}\right)\right]^{\frac{1}{2}} V_{target} \equiv \gamma\, V_{target} \quad (5)$$

Inserting the above estimates of PAC-3 MSE (interceptor) and CAV-H (target) parameters for angles-of-attack in the range of 10 to 20 degrees gives $\gamma$ in the range 0.8 to 1.2. This leads to an approximate condition for the interceptor to be able to reliably intercept the target:

$$V_{int} \gtrsim V_{target} \quad (6)$$

This condition implies that for an engagement between an interceptor like the PAC-3 MSE and a hypersonic vehicle like the HTV-2, the interceptor cannot be expected to reliably intercept vehicles traveling at speeds much greater than its own. One would therefore expect the MSE is not able to reliably intercept a hypersonic vehicle traveling faster than about Mach 6 during its dive toward the ground.

The condition given in Equation 6 will not depend strongly on details of the vehicles such as their mass and lift coefficient, because those parameters appear within a square root. This suggests that the key issue in developing a more capable endo-atmospheric interceptor is to increase its speed.

We note that the speeds typically cited for BGVs are their initial glide speeds, which will be considerably higher than their speeds at low altitudes during their dives because of drag during reentry. It is the speed at the engagement altitude that is relevant in Equation 6.

Figure 5 shows how the speed of a hypersonic vehicle with characteristics like the HTV-2 will drop during its dive, based on calculations described in Paper 1. The curves in Figure 5 start at the equilibrium glide altitude of the vehicle for initial velocities in the range Mach 5 to 10; the vehicle is then assumed to dive by rolling over and using its lift force to pull it toward the ground.

The red boxes in Figure 5 show the approximate regions of this plot where interceptors like the PAC-3 CRI and PAC-3 MSE would likely be able to intercept the vehicle during its dive, based on the analysis above. This assumes that they would not be able to intercept at very low altitudes, which could reduce the speed requirement of the interceptor by a small amount if feasible.

These results suggest that a glider like the HTV-2 traveling at about Mach 9 or faster at the start of its dive should be able to maintain speeds above Mach 6, and thus avoid interception by an interceptor like the MSE, throughout the dive. If a key role of hypersonic weapons is to evade or attack defenses, this incentivizes flying these weapons in ways that maintain their speeds above about Mach 9 throughout their glide phase.

As an example, for a glide range of 1,000 km, a glider like the HTV-2 with $L/D$ = 2.6 would need to begin its glide with a speed greater than Mach 12 since it would slow during the glide to approximately the Mach 9 threshold for evading interception. This requirement of such high speed drives up the mass of the booster required to launch a BGV. Alternately, this requirement would limit the distance that a BGV could glide and still have sufficient speed at the start of its dive.

The curves in Figure 5 assume the ballistic coefficient of the BGV during the dive phase is $\beta$ = 13,000 kg/m$^2$.[69] Note that the vehicle's value of $\beta$ during the dive may be larger than its value during glide since it may dive with a smaller angle of attack than it uses to optimize L/D during glide. For smaller values of $\beta$, the BGV will slow more during the dive, requiring it to start its dive with a speed greater than Mach 9 to evade defenses. For $\beta$ = 10,000 kg/m$^2$ it would need to start its dive with a speed greater than Mach 10; for $\beta$ = 7,000 kg/m$^2$ it would need a speed near Mach 12.[70]

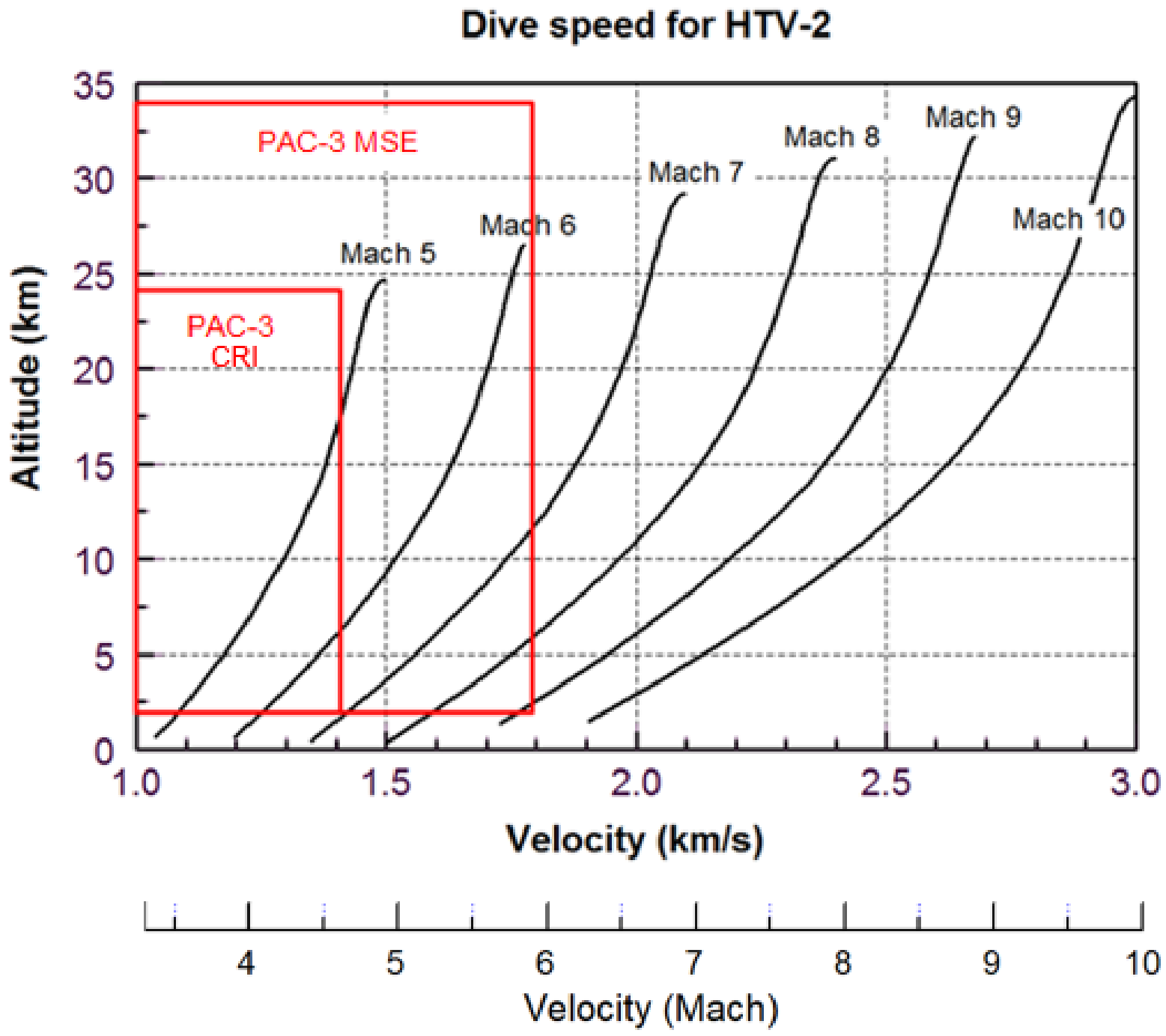

Figure 5. The speed of a vehicle similar to the HTV-2 during its dive toward the ground from its equilibrium glide altitude. Each curve is labeled by the glide speed of the vehicle at the start of its dive (i.e., at the end of its glide). The red boxes show the regions of this plot where the PAC-3 CRI and PAC-3 MSE would likely be able to intercept such a vehicle during its dive. This plot assumes $\beta$ = 13,000 kg/m$^2$ during the dive.

Interceptors like the current Aegis SM-2 and SM-6 have speeds of about Mach 4 (slower than the PAC-3 CRI) and are therefore likely less effective against maneuvering hypersonic weapons than are Patriot systems—their effective intercept region would be smaller than the red PAC-3 CRI box in Figure 5. While they have fragmentation warheads and therefore do not need to make direct hits on an HGV to disable it, they must still have small miss distances for the fragmentation warheads to effectively kill a vehicle. The planned SM-6 Block IB is said to have a speed greater than Mach 5, which would give it somewhat greater capability, but still likely less than PAC-3 MSE.[71]

While this paper does not deal explicitly with hypersonic cruise missiles, one can show that the rapid increase in drag due to the exponential increase in atmospheric density a vehicle encounters during its dive makes it impractical to use a scramjet to maintain a speed greater than Mach 6 if the vehicle starts its dive near Mach 6, since the required thrust would lead to an impractically large engine and fuel mass. The drag encountered at altitudes of 5 to 10 km would be 20 to 40 times as large as the drag encountered during glide at 30 km altitude. As a result, even powered gliders would require a high speed at the beginning of their dive to be able to evade defenses.

*Implications for Defense against BGVs*

These results suggest that defenses similar to the PAC-3 MSE should be able to engage BGVs with glide speeds at the start of dive of up to roughly Mach 9 and may therefore be able to defend small regions around high-value sites against these vehicles. Moreover, these defenses should be able to use ground-based radars to track and engage BGVs gliding at 30 to 40 km altitude, since they could detect these vehicles at about 400 km range. The performance of these radars could be aided by cueing from an external missile detection and tracking system, such as space-based early warning systems like the US Space-Based IR System (SBIRS).[72] Cueing would alert the radars to the presence and approximate location of the BGV, allowing for faster acquisition by these radars to provide precise tracking of the incoming vehicle.

On the other hand, current ship-based interceptors like the SM-2 and SM-6 may be too slow to effectively intercept such vehicles. Equipping ships with faster interceptors could therefore be a priority in defending against BGV threats.

In response to the development of BGVs, the United States is considering the deployment of constellations of space-based sensors intended to detect and track gliders over much larger areas than is feasible with ground-based radars and with better resolution than SBIRS. The proposed Hypersonic and Ballistic Tracking Space Sensor (HBTSS) satellite system is an example.[73] One goal is to use such sensors to provide tracking data accurate enough to guide interceptors to a BGV. Such tracking would appear to require sensors on a large constellation of many hundreds of satellites in low Earth orbit, in addition to satellites at higher altitudes. Deploying and maintaining such a system would be difficult and expensive.

The above analysis, however, suggests that the most important next step in improving defenses against BGVs is developing more capable interceptors, rather than developing new sensors.

This view is reinforced by an animation released by the Missile Defense Agency (MDA) that shows SM-6 interceptors and the proposed Glide Phase Interceptor (GPI) engaging hypersonic weapons attacking US ships.[74] The video indicates that the MDA believes these systems could defend ships without requiring input from a new space-based sensor network like HBTSS. While such a system would clearly offer new detection and tracking capabilities, it does not actually appear to play a crucial role in the defense.[75]

HBTSS is intended to provide sufficiently high-quality track data to allow the launch of an interceptor before the attacking vehicle is within the range of the defense system's radar; this scenario is called "engage-on-remote." However, the MDA video shows that engage-on-remote is not needed.

Instead, as the video shows, forward-based radars on ships surrounding a carrier could detect and track a BGV earlier than a radar that was near the glider's target, allowing the interceptor to be launched before the BGV was within the range of that second radar—a situation called "launch-on-remote." In addition, SBIRS may be able to cue surface-based radars. While not shown in the video, it is also possible that the defense could execute engage-on-remote using the information from the forward-based radar.

The video also shows that a capable interceptor could engage a BGV even when launched in the relatively short period after the target was detected by the interceptor's co-located radar—a situation called "cued-organic-defense." This again suggests that HBTSS could increase the area defended by interceptors but is not required to protect ships.

While the video shows the BGV ultimately being intercepted by an SM-6, the capability of the SM-6 would be limited by its speed; it would not be able to engage sufficiently fast BGVs, as discussed above.

**Conclusions**

Our analysis of current missile defense systems shows that hypersonic weapons with ranges considered here may be vulnerable to endo-atmospheric interceptors when the BGV or MaRV are in the final dive portions of their trajectories. A key result is that evading current missile defense systems requires vehicles to maintain speeds substantially higher than Mach 5 throughout their glide.

Comparing the relative maneuverability of hypersonic vehicles and current endo-atmospheric missile defense interceptors indicates that to avoid interception the speed of a reentering vehicle throughout its dive to the target should be greater than that of the interceptor. That condition sets a minimum speed that either a BGV or MaRV must maintain throughout its flight.

In particular, to evade the most capable current defense systems, a BGV or MaRV must maintain a speed greater than about Mach 6 until it reaches an altitude of a few kilometers, below which interception is not practical. For a vehicle with a ballistic coefficient during dive of $\beta$ = 13,000 kg/m$^2$, which is typical of a modern ballistic reentry vehicle, this condition implies that a BGV should begin its dive from glide phase with a speed of about Mach 9 or greater. Smaller values of $\beta$ will require the speed at the end of glide phase to be higher than Mach 9.

Because of the velocity losses of the BGV during pull-up and glide phases, a BGV requires a higher burnout speed than a MaRV to achieve the same speeds during reentry.

Moreover, the required minimum speed of a BGV will increase as new generations of faster endo-atmospheric defenses are developed.

One consequence of this analysis is that the main limitation of defending against hypersonic vehicles appears to be the speed of the interceptors of current endo-atmospheric missile defenses, and not the ability to detect and track the vehicles. Since endo-atmospheric defenses can only defend relatively small areas around the location where the interceptor is deployed, the range at which ground-based radars can begin tracking these vehicles is expected to be long enough to allow these interceptors to engage the incoming vehicles. The fact that tracking does not appear to be the primary barrier to defending against these weapons may call into question the current need for building space-based sensors systems like HBTSS.

^^^^^^^^^^^^^^^^^^^^^^^^^^^^^^^^^^^^^^^^^^^^^^^^^^^^^^^^^^^^^^^^^^^^^^^^^^^^^^^^^^^^^^^^^^^^

## Appendix A: Implications of Increasing *L*/*D*

A key parameter governing the capabilities of a hypersonic weapon is the lift-to-drag ratio, *L*/*D*. While *L/D* for a subsonic aircraft can be in the range of 15 to 20, the values for hypersonic gliders are much lower; as noted in the text, *L/D* for the BGVs the United States is developing appear to be less than three.

It is useful to ask to what extent future BGVs might use designs and materials that could significantly increase *L/D*, and to what extent increasing *L/D* might significantly improve the military capabilities of these weapons.

Studies show that in principle it may be possible to increase the *L*/*D* values of hypersonic vehicles to six or higher using waverider designs described in the text.[76] Increasing *L*/*D* to four or six would increase the maximum range of an HGV at a given glide speed and therefore help to reduce the amount of energy needed to reach a given range. This could reduce the mass of air-launched BGVs, and could allow HCMs to carry less fuel for a given range. For many applications, simply using a booster large enough to achieve the desired range is likely preferable to the difficult task of increasing range by increasing glider L/D.

However, while the waverider concept dates from the late 1950s it has proved difficult to turn into working vehicles with high *L/D*. As a result, *L/D* of future systems may increase slowly over current values.

As we show, the effects of realistic increases in L/D on militarily relevant capabilities of BGVs are generally minor.

*Effects of Changing L/D and β*

The parameters that determine the most important aerodynamic properties of a vehicle are its lift and drag coefficients, $C_L$ and $C_D$, respectively. These are typically discussed in terms of the lift-to-drag ratio, $L/D = C_L/C_D$, and ballistic coefficient, $\beta = m/(C_D A)$, where $A$ is a reference area of the vehicle.

The dynamics of the vehicle during glide (velocity profile, range, and flight time) depend primarily on $L/D$ and very weakly on $\beta$ for a given initial glide speed $V$ and vehicle mass $m$. This can be seen from Equation 18 in the main text and the definition of $L/D$, which give:

$$F_L = \frac{1}{2} C_L A \rho V^2 = \alpha(V) m g \quad (A1)$$

$$F_D = \frac{1}{2} C_D A \rho V^2 = \frac{\alpha(V) m g}{L/D} \quad (A2)$$

Where $\alpha(V) = 1 - (V/V_e)^2$. These equations show that the forces acting on the vehicle during its glide only depend on $\beta$ through the small effect the equilibrium glide altitude, which depends on $\beta$, has on the gravitational potential.[77] As a result, the primary effect of changing $\beta$ will be to change the drag at a given altitude, but this will also change the glide altitude in a way that the vehicle feels essentially the same forces. This implies that determining the parameters of the HTV-2 from its test data will constrain the value of L/D more tightly than the value of $\beta$.[78]

Equation 26 shows that the range of a BGV increases with $L/D$. However, since the range depends on the square of the glide speed, that range increase could alternately be achieved by a small increase in $V$.

Heating of HGVs during the glide phase, which is a key issue for their development, depends on both $L/D$ and $\beta$, and we next consider that dependence.

The glide altitude $h$ can be determined from the density at the glide altitude, which from Equation A1 depends on the lift coefficient and can be written:

$$\rho(h) = \frac{2\alpha g\beta}{{}^{L}\!/_{D}\, V^2} \qquad (A3)$$

Because of Equation A2, the change in the vehicle's kinetic energy due to drag, which shows up as energy transfer to the air, depends on *L/D* but only very weakly on $\beta$. In particular, the rate of change of kinetic energy (KE) (ignoring the small change in potential energy during glide) equals the drag force times $V$, and using Equation A2 is:

$$\frac{dKE}{dt} = F_D V = \frac{\alpha m g V}{{}^{L}\!/_{D}} \qquad (A4)$$

Ignoring the minor velocity dependence of $\alpha = (1-V^2/V_e^2)$ for speeds less than about Mach 12, this gives:

$$\Delta KE = \frac{\alpha m g}{{}^{L}\!/_{D}} \int V dt = \frac{\alpha m g}{{}^{L}\!/_{D}}\, r_G \qquad (A5)$$

so that the energy lost by the vehicle for a given glide range is inversely proportional to *L/D*. However, since the glide range for a BGV starting at $V_g$ and ending at $V_f$ is proportional to *L/D*, the total energy loss of the vehicle in that case is independent of *L/D* and depends only on the speeds at the start and end of glide, as expected:[79]

$$\Delta KE = \frac{\alpha m g}{{}^{L}\!/_{D}}\, r_G = \frac{m}{2}\left(V_g^2 - V_f^2\right) \qquad (A6)$$

where the last equality uses Equation 26 for the glide distance $r_G(V_g, V_f)$.

Only a fraction of the energy lost by the BGV, however, is transferred from the air to the vehicle and results in heating of the body. A commonly used measure of the energy transfer to the vehicle is $dq/dt = \rho V^3$ (note that the empirical heating equations used in Paper 1 depend roughly on $\rho V^3$) where $q$ is the energy absorbed per area. Equation A3 gives:

$$\frac{dq}{dt} = \rho V^3 = \frac{2\alpha g V \beta}{L/D} \qquad (A7)$$

Unlike Equation A4, this quantity depends on $\beta$ because the fraction of energy that is transferred from the air to the vehicle depends on the atmospheric density around the body, and the density at the glide altitude depends on $\beta$. Unlike the dynamics if the vehicle, the heating will therefore depend on both $\beta$ and *L/D*.

Since the vehicle's surface temperature $T$ is proportional to the fourth root of *dq/dt* from the Stefan-Boltzmann law, the body temperature changes slowly with $\beta/(L/D)$.[80] Numerical calculations using the methods described in Paper 1 show that $T$ is roughly proportional to $[\beta/(L/D)]^a$, where $a$ = 1/7 to 1/5 for speeds below Mach 12, confirming that $T$ varies only slowly with $\beta/(L/D)$.

For example, decreasing $\beta/(L/D)$ by a factor two, which would correspond, e.g., to increasing *L/D* from the HTV-2 value 2.6 to 5.2 (at constant $\beta$), would only reduce a glider's surface temperature by about 10 percent.

Moreover, if the vehicle takes advantage of the longer glide range that is possible with higher *L/D*, then the total energy absorbed by the vehicle would depend on $\beta$ but not *L/D.* Ignoring the minor velocity dependence of $\alpha$ for speeds less than about Mach 12, one finds:

$$q = \frac{2\alpha g \beta}{L/D} \int V dt = \frac{2\alpha g \beta}{L/D} r_G = \beta\left(V_g^2 - V_f^2\right) \qquad (A8)$$

where $r_G$ is the glide distance of the BGV starting at $V_g$ and ending at $V_f$, which is proportional to *L/D* (Equation 26). The fact that $q$ is proportional to $\beta$ may be surprising since it means the energy absorbed decreases with a larger drag coefficient, but this is essentially the reason that early heat shields for spacecraft and reentry vehicles were blunt, which allowed these vehicles to slow at altitudes where atmospheric density was low.

For a given *L/D* and $\beta$, an effective way of limiting the heat load to a BGV is to limit its speed (which decreases the heat transfer rate) and/or to shorten its range (which reduces the duration of heating). This appears to be the current US approach, since it is focusing on gliders with speeds below about Mach 12 and ranges of only up to a couple thousand kilometers, rather than long-range vehicles like the HTV-2. Reduced heating of a BGV would also decrease its infrared radiant intensity and therefore its visibility to IR sensors, such as SBIRS.

**Appendix B: Coordinate System Used in Paper 1**

Our previous paper uses simplified equations of motion (Equations (1) – (6) in that article) that are appropriate for the quantities we calculated.[81] In the corresponding coordinate system, $\Psi$ and $\Omega$ are the down-range and cross-range angles, with distances measured along great circles so that $\Psi r_e$ is the range and $\Omega r_e$ is the cross-range distance.[82] The advantage of these coordinates for our calculation is that the velocity angle $\kappa$ has a simple physical interpretation as the direction of motion relative to the down-range direction.

Also, as we discuss below, using the full set of equations in spherical coordinates requires an additional set of equations to determine what an observer on the Earth would consider range and cross-range distances, due to the nature of spherical coordinates near the poles.

The full set of equations of motion in spherical coordinates (ignoring Earth's rotation), as given in reference 21 of the original paper, would replace Equations (3) - (5) in our paper with these equations:

$$\frac{d\eta}{dt} = \left({}^{L}/_{D}\right)\left(\frac{C_d A}{2m}\right)\frac{\rho v \sin\sigma}{\cos\gamma} - \frac{v\, tan\theta\, \cos\gamma \sin\eta}{r_e + h} \quad (3')$$

$$\frac{d\theta}{dt} = \frac{v \cos\gamma \cos\eta}{r_e + h} \quad (4')$$

$$\frac{d\phi}{dt} = \frac{v \cos\gamma \sin\eta}{(r_e + h)\cos\theta} \quad (5')$$

where $\theta$ and $\phi$ are geographic latitude and longitude coordinates, and $\eta$ is the heading angle of the velocity vector measured from north. The origin of the coordinate system, and the launch location of the vehicle, is a point on the equator with zero latitude and longitude. The vehicle is assumed to be launched north along the line of zero longitude.

These equations differ from (3) and (5) by the appearance of the second term in (3') and the factor of $\cos\theta$ in (5'). (The appearance of radius $r_e$ rather than $r = r_e+h$ in Equations (4) and (5) was a typo in the original paper; our calculations used the correct version.)

Using Equations (3') - (5') requires new equations for down-range and cross-range distances. These distances (measured along great circles) for a vehicle at a point with latitude and longitude ($\theta$,$\phi$) are given by:

$$crossrange = \varepsilon r_e \quad (B1)$$

$$range = r_e \, cos^{-1}\left(\frac{\cos\theta\cos\phi}{\cos\varepsilon}\right) \quad (B2)$$

where

$$cos^2\,\varepsilon = (1 + sin^2\theta + cos^2\theta\cos(2\phi))\,/2 \quad (B3)$$

The geometrical difference between the two coordinate systems and a derivation of Equations (B1) – (B3) is discussed below.

Both sets of equations give identical results for flight purely in the down-range direction. They also give differences of only a few percent or less in range and cross-range for the maneuvering calculations presented in the paper, the results of which are shown in Figures 6 and 14 of that paper.

*Calculating Range and Cross-range in Different Coordinate Systems*

Assume a vehicle begins flying at the origin and at some time is located at a point *P*. Figure B1 shows in blue how *P*'s coordinates $(\theta, \phi)$ are defined in terms of latitude and longitude. Flight begins in the north direction, so the longitude = 0 circle defines the range direction.

It is important to note that given a point *P*, there are (at least) two ways to associate a point on the longitude = 0 line with point *P*, and to define the "cross-range" distance between *P* and the longitude = 0 line:

(1) Draw a latitude circle through *P* (that circle is defined by the intersection of the sphere with a plane parallel to the equatorial plane). Define point *A* as where that latitude line crosses the longitude = 0 line (see Figure B1). θ is the angle between the x-axis and a radial line from the center of the sphere to point *A*. In this case, the distance from *P* to *A* along the latitude line is $\phi\, Re \cos\theta$, but that is not what you would call the "distance between *P* and *A*" which would instead be defined along a geodesic of the sphere, that is, along a great circle.

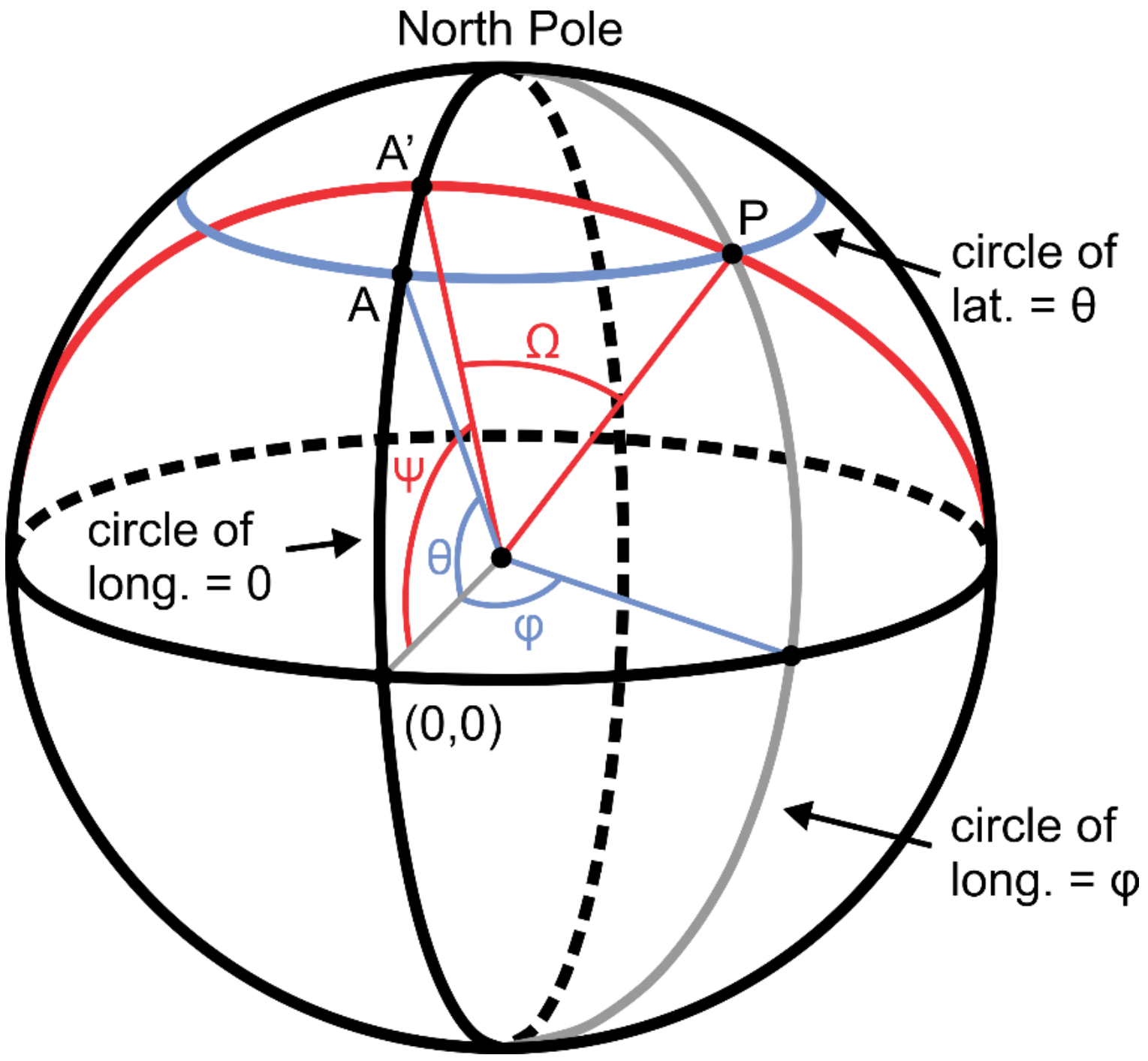

**Figure B1.** This figure shows two ways of assigning coordinates to a point *P.* Standard spherical coordinates for *P* are shown in blue. The vehicle is launched at (0,0) and the "range" direction is along the circle of longitude = 0 toward the north pole. The "cross-range" distance to *P* is the distance measured along the red great circle to *A'.*

(2) The correct way to associate range and cross-range distances with *P* is to draw a great circle through *P* that is perpendicular to the longitude = 0 line; this is shown in red in Figure B1. It crosses the longitude = 0 line at the point *A'*. The distance from the origin to *A'* is what an observer on the Earth would define as the "range" of a vehicle that had reached *P*, since the "cross-range" distance to *P* would be measured along a great circle passing through *A'* and *P*. The associated angles in this case, $\Psi$ and $\Omega$, are shown in red.

Figure B1 shows that $\Psi$ is always larger than $\theta$, although near the equator the difference is small.

Figure B2 illustrates the difference between $\theta$ and $\Psi$ in an illustrative case. This figure is looking down on the north pole; the vehicle is launched from the equator and is passing by the pole on the red trajectory. It is clear that what we would call the "range angle" ($\Psi$) is 90 degrees, but in this case the spherical-coordinate angle $\theta$ never gets larger than 85 degrees.

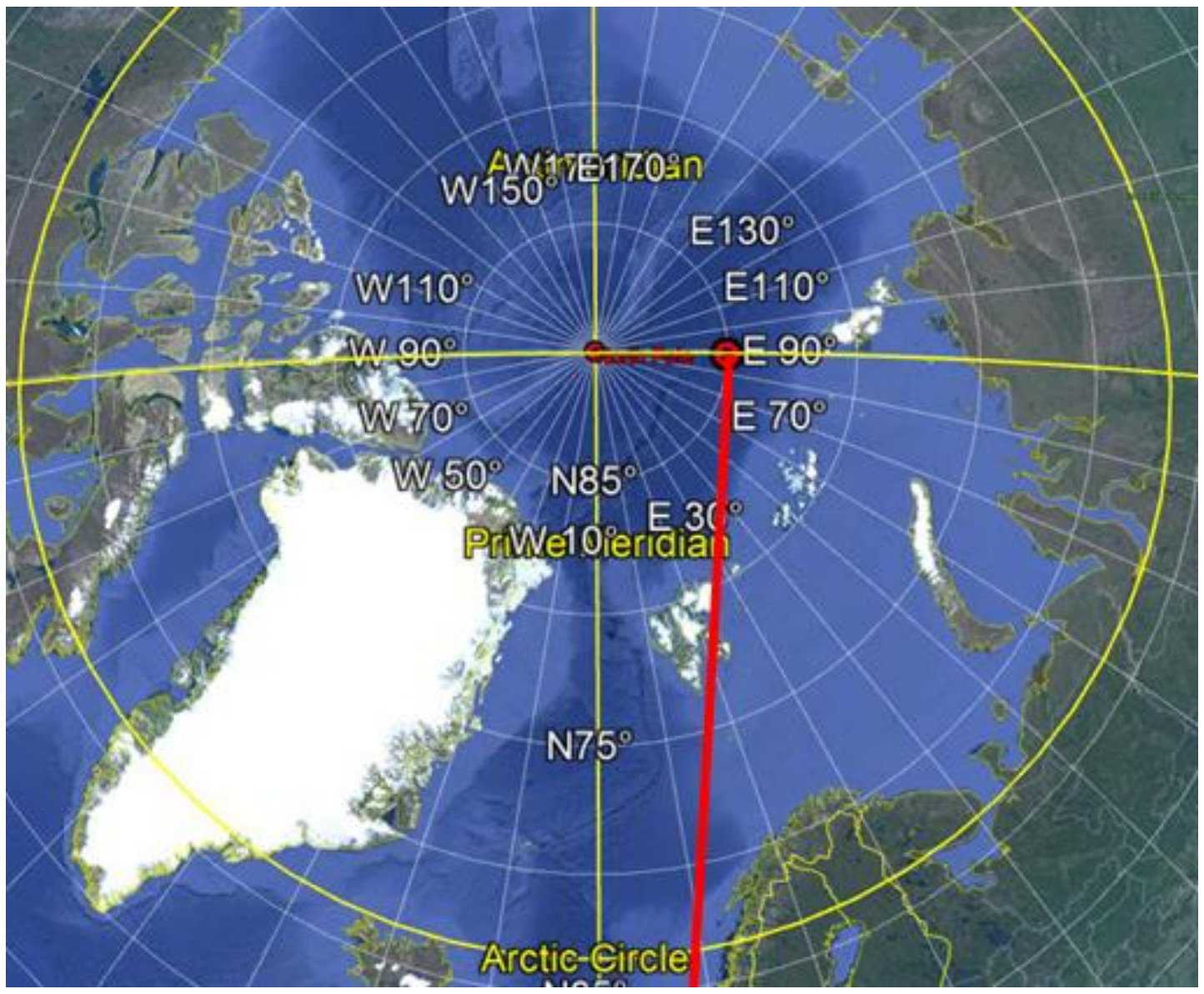

**Figure B2.** This figure shows a view from above the north pole of a vehicle launched from the origin; the trajectory is shown in red.

*Derivation of Equations for Range and Cross-range from Spherical Coordinates*

Figure B1 shows that the spherical coordinates angles $\theta$ and $\phi$ do not directly give the range and cross-range distances to a point *P*. This section derives equations for range and cross-range from those variables.

Consider a point at (latitude, longitude) = $(\theta, \phi)$ (see Figure B3).

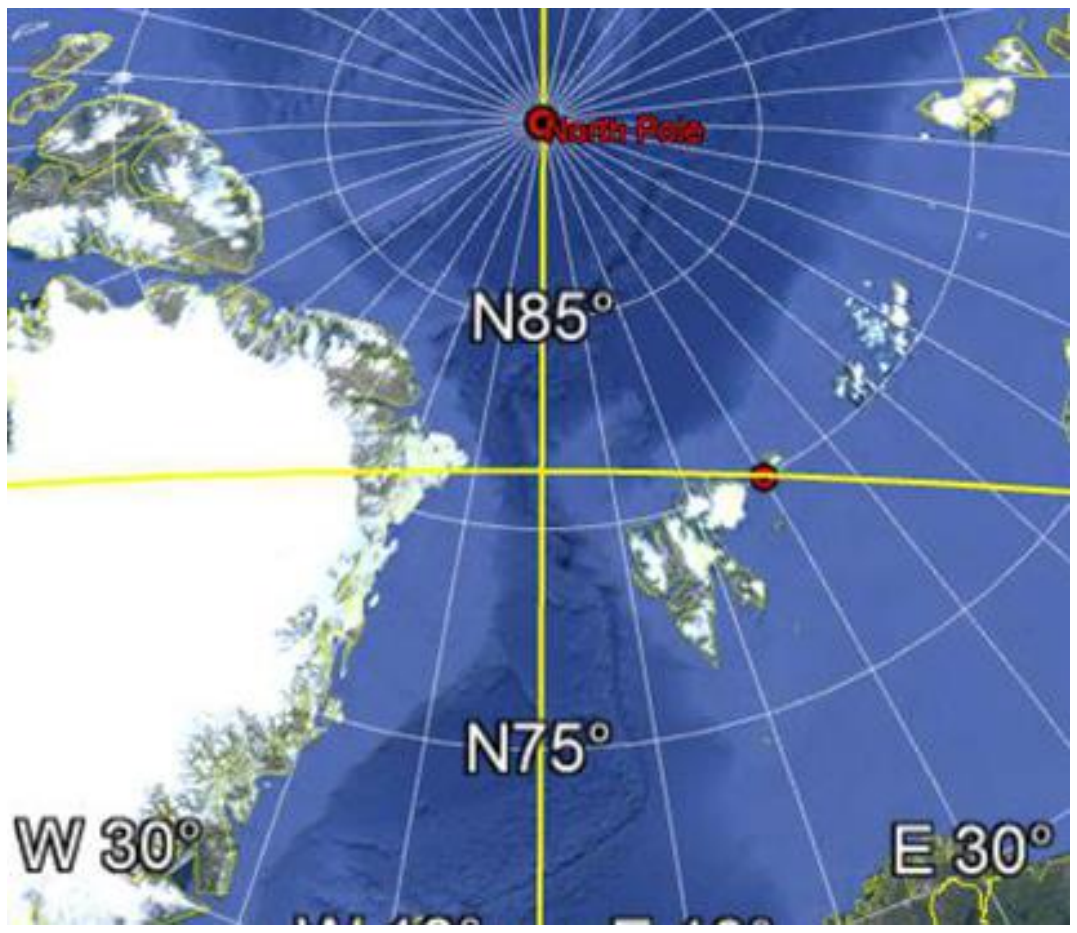

**Figure B3:** Consider the point shown in red at (N80, E30). The cross-range distance to the point is measured from the latitude = 0 line along a great circle, show as the horizontal yellow curve. Note that the distance measured along a line of latitude is a fraction $\phi/2\pi$ of the circumference of a circle at latitude $\theta$, which is $2\pi\ r_e \cos\ \theta$. So this distance is $\phi\ r_e \cos\theta$.

The cross-range distance is measured along a great circle through the point, and can be found using spherical geometry:

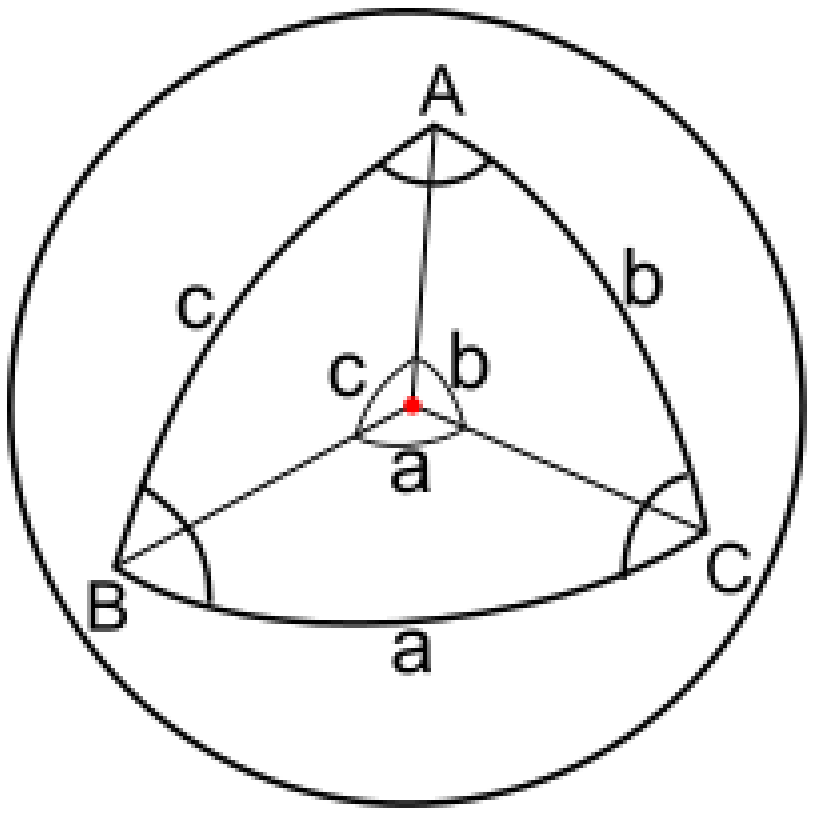

**Figure B4:** This figure assumes a unit sphere. In general the length of the arc subtended by the angle *a* will be *ar*, where *r* is the radius of the sphere. Here *A* is the angle between the arcs *AB* and *AC*, and *a* is the angle in the horizontal plane containing points *B* and *C*.

Consider a spherical triangle, with sides made of segments of great circles, as in Fig B4. Take *A* to be at the north pole, and *B* and *C* to be the points $(\theta, \pm\phi)$. The angle $a$ is $2\phi$, and angles $b$ and $c$ are $(90 - \theta)$.

The spherical law of cosines:[83]

$$\cos a = \cos b \cos c + \sin b \sin c \cos A \qquad (B5)$$

gives:

$$\cos a = sin^2\theta + cos^2\theta\cos(2\phi) \qquad (B6)$$

The cross-range is half the length of the great circle segment connecting b and c, which can be found from Equation (C6) as:

$$Crossrange = \frac{a}{2} r_e \qquad (B7)$$

The range angle corresponding to a point with (latitude, longitude) = $(\theta,\phi)$ can be found in a similar way. Consider the line of longitude = 0, which bisects the angle *A* and the length *a* in Figure B4; this line is shown in red in Figure B5, which is the black triangle is that of Figure B4 with the points relabeled. The point where this line intersects the horizontal arc is labelled *F*. The range angle will be given by 90 – *e*, where *e* is the angle between the radial vectors from the center of the sphere to the points D and F.

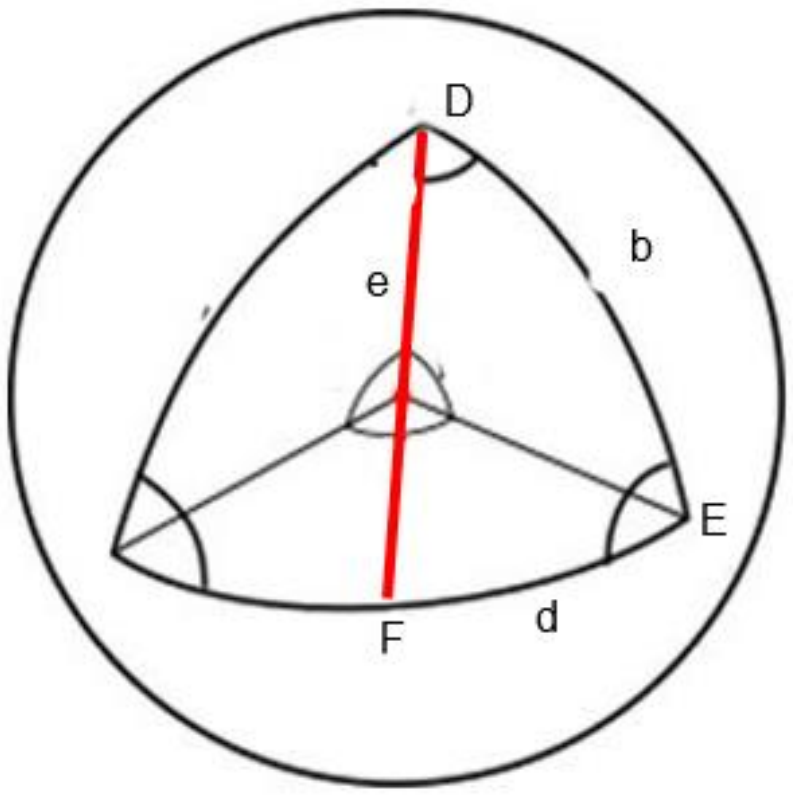

**Figure B5.** This figure shows the triangle from Figure B4 with the point relabeled. The red arc connecting *D* and *F* bisects the angle at *A* in the previous figure.

The angle $D = A/2 = \phi$, and the length $d = a/2\ r_e$, where $a$ was calculated above.

From the spherical law of sines:[84]

$$\frac{\sin D}{\sin d} = \frac{\sin E}{\sin e} \qquad (B8)$$

so that

$$\sin e = \sin E\ \frac{\sin d}{\sin D} \qquad (B9)$$

From Figure B5:

$$\frac{\sin A}{\sin a} = \frac{\sin C}{\sin c} = \frac{\sin E}{\sin c} \qquad (B10)$$

where the last equality comes from the fact that angles *C* and *E* are the same in the two cases.

Combining these gives:

$$\sin e = \left(\frac{\sin\frac{a}{2}}{\sin a}\right)\left(\frac{\sin 2\phi}{\sin\phi}\right)\cos\Psi = \frac{\cos\phi\cos\theta}{\cos\frac{a}{2}} \qquad (B11)$$

or

$$\sin e = \frac{\cos\phi\cos\theta}{\left(\frac{1+\cos a}{2}\right)^{1/2}} \qquad (B12)$$

where *cos a* is given by Equation (B6). The range is then given by:

$$Range = \left(\frac{\pi}{2} - e\right) r_e \qquad (B13)$$

So for a point with (latitude, longitude) = *(θ, ϕ),* the range and cross-range distances are given by Equations (B7) and (B13).

Letting *ε* = *a/2*, so that it is the cross-range angle, these equations can be written in the alternate form given in Equations (B1) - (B3).

---

[1] Visiting Scholar, Laboratory for Nuclear Security and Policy, Department of Nuclear Science and Engineering, MIT

[2] Research Scholar, Center for International Security and Cooperation (CISAC), Freeman Spogli Institute for International Studies, Stanford University

[3] The speed of sound in the atmosphere varies by about 10 percent over the range of altitudes of interest for hypersonic weapons (10 to 50 km). Throughout this paper we assume a speed of 300 m/s, which is roughly consistent with a standard engineering approximation that uses 1000 ft/s as sound speed at these altitudes. See "1976 Standard Atmosphere Calculator," DigitalDutch, https://www.digitaldutch.com/atmoscalc/table.htm

[4] Richard H. Speier, George Nacouzi, Carrie A. Lee, and Richard M. Moore, *Hypersonic Missile Nonproliferation: Hindering the Spread of a New Class of Weapons* (Santa Monica, CA: RAND Corporation, 2017), 53-93, https://rand.org/pubs/research_reports/RR2137.html

[5] MaRVs were developed and tested during the Cold War and in the 2000s. See Matthew Bunn, "Technology of Ballistic Missile Reentry Vehicles," in Review of U.S. Military Research and Development: 1984, eds. Kosta Tsipis and Penny Janeway (Mclean, VA: Pergamon, 1984), 87–107, https://scholar.harvard.edu/files/bunn_tech_of_ballistic_missle_reentry_vehicles.pdf ; Amy Woolf, *Conventional Prompt Global Strike and Long-Range Ballistic Missiles: Background and Issues* (Washington, DC: Congressional Research Service, 2021), https://crsreports.congress.gov/product/pdf/R/R41464

[6] James M. Acton, "Hypersonic Boost-Glide Weapons," *Science and Global Security* 23 (2015): 191–219, http://scienceandglobalsecurity.org/archive/sgs23acton.pdf ; David Wright, "Research Note to Hypersonic Boost-

Glide Weapons by James M. Acton: Analysis of the Boost Phase of the HTV-2 Hypersonic Glider Tests," *Science and Global Security* 23 (2015): 220–229 http://scienceandglobalsecurity.org/archive/sgs23wright.pdf

[7] Cameron L. Tracy and David Wright, "Modelling the Performance of Hypersonic Boost-Glide Missiles," *Science and Global Security* 28 (2021): 135-170, http://scienceandglobalsecurity.org/archive/sgs28tracy.pdf In this paper we use a different coordinate system for the equations of motion than is used in many papers. That system and the reasons behind it are described in Appendix B.

[8] "The First Missile Regiment of Avangard Took Up Combat Duty," *TASS*, 27 December 2019, https://tass.ru/armiya-i-opk/7436431 ; "Deployment of Avangard continues in Dombarovskiy," Russian Strategic Nuclear Forces, 16 December 2020, https://russianforces.org/blog/2020/12/deployment_of_avangard_continu.shtml

[9] "Avangard," Center for Strategic and International Studies, 31 July 2021, https://missilethreat.csis.org/missile/avangard/

[10] An additional motivation not listed here appears to be a desire to match work on hypersonic weapons by Russia and China. See, for example, Oren Liebermann, "US is Increasing Pace of Hypersonic Weapons Development to Chase China and Russia, Senior Admiral Says," *CNN*, 20 November 2022, https://www.cnn.com/2022/11/20/politics/us-hypersonic-china-russia-competition/index.html.

[11] Ivett A. Leyva, "The Relentless Pursuit of Hypersonic Flight," *Physics Today*, 70 (2017): 30-36, https://doi.org/10.1063/PT.3.3762

[12] See, for example, Joseph Trevithick, "Here's How Hypersonic Weapons Could Completely Change the Face of Warfare," *The War Zone*, 6 June 2017, https://www.thedrive.com/the-war-zone/11177/heres-how-hypersonic-weapons-could-completely-change-the-face-of-warfare ; Benjamin Knudsen, "An Examination of U.S. Hypersonic Weapon Systems," Technical Report, George Washington University, June 2017, DOI:10.13140/RG.2.2.14375.96164 ; "Hypersonic Strike and Defense: A Conversation with Mike White," *Center for Strategic and International Studies,* 10 June 2021, https://www.csis.org/analysis/hypersonic-strike-and-defense-conversation-mike-white . Another possible motivation, reducing warning of an attack by avoiding detection, does not appear to be as central as these. Ground-based radars will not see an HGV flying at low altitudes until it is within less than 500 km, but this may not be particularly relevant. The detection range would likely still be long enough to provide time to launch short-range interceptors against it. For high-value targets, forward-basing radars would provide additional warning time. Moreover, US and Russian space-based infrared (IR) sensors (which China is currently developing) can detect and provide early warning of the launch of the boosters carrying hypersonic weapons. In addition, these sensors will likely be able to detect the IR emissions from these weapons when they are gliding at sufficiently high speeds, providing warning and cuing information even if this data is not sufficient to guide interceptors (see Paper 1).

[13] This paper focuses on terminal, endo-atmospheric defenses for several reasons, which we discuss in detail below. While underflying midcourse (exo-atmospheric) defenses may be a motivation for developing long-range HGVs, the current focus is on shorter range HGVs that would be difficult to attack with exo-atmospheric defenses. In addition, countermeasures that could be used to defeat midcourse defenses are well-known and could accompany boosters

used to launch either MaRVs or HGVs. Moreover, no country other than the United States has deployed modern exo-atmospheric defenses, so they are not currently relevant to US motivations for HGV development.

[14] See Paper 1 and analysis below.

[15] Acton, "Hypersonic Boost-Glide Weapons."

[16] Changing $\beta$ will change estimates of heating during glide—see Appendix A and Graham V. Candler and Ivett A. Leyva, "Computational Fluid Dynamics Analysis of the Infrared Emission From a Generic Hypersonic Glide Vehicle," *Science and Global Security,* 7 December 2022, https://doi.org/10.1080/08929882.2022.2145777

[17] Acton, "Hypersonic Boost-Glide Weapons."

[18] "Minimum-energy trajectories" give the maximum range for a given burnout speed and altitude.

[19] Lisbeth Gronlund and David C. Wright "Depressed-Trajectory SLBMs: A Technical Assessment and Arms Control Possibilities," *Science and Global Security* 3 (1992): 101-159, http://scienceandglobalsecurity.org/archive/sgs03gronlund.pdf

[20] While Russia is developing short-range HGVs, the Kinzhal system that it reportedly used against Ukraine in 2022 is described as a maneuvering air-launched ballistic missile rather than an HGV (Kelley Sayler, *Hypersonic Weapons: Background and Issues for Congress* (Washington, DC: Congressional Research Service, 2022), https://crsreports.congress.gov/product/pdf/R/R45811/25 )). The "hypersonic weapon" North Korea has reportedly tested is also described this way (Iain Marlow and Jon Herskovitz, "Kim Jong Un's Hypersonic Missiles Show He Can Hit U.S. Back," *Bloomberg*, 12 January 2022, https://www.bloomberg.com/news/articles/2022-01-12/kim-jong-un-s-new-hypersonic-missiles-show-he-can-hit-u-s-back ).

[21] Richard Hallion, "The History of Hypersonics: or, 'Back to the Future—Again and Again'," American Institute of Aeronautics and Astronautics, AIAA-2005-329 (2005), 43rd AIAA Aerospace Sciences Meeting and Exhibit, 10-13 January 2005, Reno, NV, https://doi.org/10.2514/6.2005-329

[22] Woolf, *Conventional Prompt Global Strike.*

[23] "An Historical Overview of Waverider Evolution," Staar Research, https://www.gbnet.net/orgs/staar/wavehist.html

[24] To address this issue, research is currently being done on HGVs designed to optimize lift at more than one velocity; see, e.g., Zhen-tao Zhao, Wei Huang, Li Yan, Yan-guang Yang, "Overview of Wide-speed Range Waveriders," *Progress in Aerospace Sciences* (2020) 113: 1-14, https://doi.org/10.1016/j.paerosci.2020.100606

[25] Acton, "Hypersonic Boost-Glide Weapons."

[26] Woolf, *Conventional Prompt Global Strike.*

[27] Kenneth W. lliff and Mary F. Shafer, "A Comparison of Hypersonic Flight and Prediction Results," *American Institute of Aeronautics and Astronautics,* AIAA-93-0311 (1993), 31st Aerospace Sciences Meeting and Exhibit, 11-14 January 1993, Reno, NV, https://doi.org/10.2514/6.1993-311

[28] Woolf, *Conventional Prompt Global Strike:* 17.

[29] "Advanced Hypersonic Weapon," *Army Technology*, 10 April 2012, https://www.army-technology.com/projects/advanced-hypersonic-weapon-ahw/ The Army also reportedly tested a "downscaled

version" in October 2017 (Sydney J. Freedberg, Jr., "Army Warhead Is Key to Joint Hypersonics," *Breaking Defense*, 22 August 2018, https://breakingdefense.com/2018/08/army-warhead-is-key-to-joint-hypersonics/ ).

[30] Acton, "Hypersonic Boost-Glide Weapons."

[31] Joseph Trevithick, "USAF, Army, and Navy Join Forces to Field America's First Operational Hypersonic Weapon," *The Drive*, 11 October 2018, https://www.thedrive.com/the-war-zone/24181/usaf-army-and-navy-join-forces-to-field-americas-first-operational-hypersonic-weapon

[32] John A. Tirpak, "Air Force Cancels HCSW Hypersonic Missile in Favor of ARRW," *Air Force Magazine,* 10 February 2020, https://www.airforcemag.com/air-force-cancels-hcsw-hypersonic-missile-in-favor-of-arrw/

[33] Yi Feng, Shenshen Liu, Wei Tang, Yewei Gui, "Aerodynamic Configuration Design and Optimization for Hypersonic Vehicles, *American Institute of Aeronautics and Astronautics* (2017), 21st AIAA International Space Planes and Hypersonics Technologies Conference, 6-9 March 2017, Xiamen, China, https://doi.org/10.2514/6.2017-2173 ; The analysis in Appendix G of National Research Council, *U.S. Conventional Prompt Global Strike: Issues for 2008 and Beyond,* Committee on Conventional Prompt Global Strike Capability (2008): 206-215, https://doi.org/10.17226/12061 assumes L/D = 2.2 for its calculations.

[34] Joseph Trevithick, "The Army and Navy Have Conducted the First Joint Test of Their New Hypersonic Weapon," *The Drive*, 20 March 2020, https://www.thedrive.com/the-war-zone/32667/the-army-and-navy-have-conducted-the-first-joint-test-of-their-new-hypersonic-weapon , Caleb Larson, "This U.S. Missile Can Kill Any Target on the Planet (In Less Than an Hour)," *National Interest*, 23 June 2020, https://nationalinterest.org/blog/buzz/us-missile-can-kill-any-target-planet-less-hour-163303 .

[35] John A. Tirpak, "Roper: The ARRW Hypersonic Missile Better Option for USAF," *Air Force Magazine*, 2 March 2020, https://www.airforcemag.com/arrw-beat-hcsw-because-its-smaller-better-for-usaf/ .

[36] Guy Norris, "High-Speed Strike Weapon to Build on X-51 Flight," *Aviation Week & Space Technology*, 20 May 2013, https://web.archive.org/web/20140104023933/http://www.aviationweek.com/Article/PrintArticle.aspx?id=%2Farticle-xml%2FAW_05_20_2013_p24-579062.xml&p=1&printView=true

[37] "X-51A Waverider," U.S. Air Force Factsheet, 3 May 2013, https://web.archive.org/web/20130619105330/http:/www.af.mil/information/factsheets/factsheet.asp?fsID=17986 ; "Hyper-X Program," *NASA* Factsheet, 28 February 2014, https://www.nasa.gov/centers/armstrong/news/FactSheets/FS-040-DFRC.html

[38] Kristen N. Roberts, *Analysis and Design of a Hypersonic Scramjet Engine with a Starting Mach Number of 4.00*, Masters thesis in Aerospace Egineering, University of Texas at Arlington, 2008, http://hdl.handle.net/10106/1073

[39] Sydney J. Freedberg, Jr., "Hypersonics: DoD Wants 'Hundreds of Weapons' ASAP," *Breaking Defense*, 24 April 2020, https://breakingdefense.com/2020/04/hypersonics-dod-wants-hundreds-of-weapons-asap/

[40] Sayler, "Hypersonic Weapons."

[41] "Kh-47M2 Kinzhal," *Missile Threat,* 19 March 2022, https://missilethreat.csis.org/missile/kinzhal/#easy-footnote-bottom-10-3801 . The Kinzhal mass is estimated at 3,800 kg (Vladimir Karnozov, "Putin Unveils Kinzhal

Hypersonic Missile," *AIN Online*, March 2, 2018, https://www.ainonline.com/aviation-news/defense/2018-03-02/putin-unveils-kinzhal-hypersonic-missile)

[42] Franz-Stefan Gady, "China Tests New Weapon Capable of Breaching US Missile Defense Systems," *The Diplomat*, 28 April 2016, https://thediplomat.com/2016/04/china-tests-new-weapon-capable-of-breaching-u-s-missile-defense-systems/ ; Sayler, "Hypersonic Weapons."

[43] Mike Yeo, "China unveils drones, missiles and hypersonic glide vehicle at military parade, *Defense News,* 1 October 2019, https://www.defensenews.com/global/asia-pacific/2019/10/01/china-unveils-drones-missiles-and-hypersonic-glide-vehicle-at-military-parade/

[44] Zhao Lei, "Superfast aircraft test a 'success'," *China Daily*, 6 August 2018, http://usa.chinadaily.com.cn/a/201808/06/WS5b6787b4a3100d951b8c8ae6.html

[45] Sayler, "Hypersonic Weapons."

[46] These values assume $\beta \sim 10{,}000$ kg/m$^2$ and vary slowly with $\beta$.

[47] Graham Warwick, "DARPA's HTV-2 Didn't Phone Home," *Aviation Week Network*, 24 April 2010, https://web.archive.org/web/20111117084740/http://www.aviationweek.com/aw/blogs/defense/index.jsp?plckController=Blog&plckBlogPage=BlogViewPost&newspaperUserId=27ec4a53-dcc8-42d0-bd3a-01329aef79a7&plckPostId=Blog%3a27ec4a53-dcc8-42d0-bd3a-01329aef79a7Post%3a70769585-4348-4701-889a-f02c58f38314&plckScript=blogScript&plckElementId=blogDest ; Ian Sample, "Falcon HTV-2 is Lost During Bid to Become Fastest Ever Plane," *The Guardian*, 11 August 2011, https://www.theguardian.com/world/2011/aug/11/fastest-ever-plane-lost-during-test-flight

[48] Thomas Newdick, "This Is Our First View Of Russia's New S-500 Air Defense System In Action," *The Drive*, 20 July 2021, https://www.thedrive.com/the-war-zone/41627/this-is-our-first-view-of-russias-new-s-500-air-defense-system-in-action

[49] Andrew M. Sessler, et al., *Countermeasures: The Operational Effectiveness of the Planned US National Missile Defense System* (Cambridge, MA: Union of Concerned Scientists and MIT Security Studies Program, April 2000): 28, https://www.ucsusa.org/sites/default/files/2019-09/countermeasures.pdf

[50] Theodore A. Postol and George N. Lewis, "The Illusion of Missile Defense: Why THAAD Will Not Protect South Korea," *Global Asia,* Vol. 11 (3), September 2016: 80-85, https://www.globalasia.org/data/file/articles/78a89c3da89bc3fae2f1e8249871c58e.pdf

[51] "Theater High Altitude Area Defense (THAAD)," Aerojet Rocketdyne, 13 March 2019, https://rocket.com/defense/missile-defense/thaad

[52] Sessler, et al., *Countermeasures.*

[53] For a discussion of acceleration saturation effects for various types of guidance, with and without lags, see Paul Zarchan, *Tactical and Strategic Missile Guidance - An Introduction (7th Edition),* (Reston, VA: American Institute of Aeronautics and Astronautics, Inc., 2019), Vol.1: 157-160, 206-211, 216-223, 254; Vol. 2: 165-167, 552-553. See also N.F. Palumbo, R.A. Blauwkamp, and J.M. Lloyd, "Modern Homing Missile Guidance Theory and Techniques," *Johns Hopkins APL Technical Digest*, Vol. 29, No. 1 (2010), 42-59, https://www.jhuapl.edu/Content/techdigest/pdf/V29-N01/29-01-Palumbo_Homing.pdf

[54] Zarchan, *Tactical and Strategic Missile Guidance*, Vol. 1: 152; Vol. 2: 147, 307, 439.

[55] Zarchan, *Tactical and Strategic Missile Guidance, Vol. 1*: 157.

[56] Terry H. Phillips, "A Common Aero Vehicle (CAV) Model, Description, and Employment Guide," Schafer Corporation (27 January 2003).

[57] Jon Hawkes "Patriot games: Raytheon's Air-Defence System Continues to Proliferate," *Jane's International Defence Review*, January 2019, Vol 52, 1-6 https://web.archive.org/web/20190601000000*/https://www.raytheon.com/sites/default/files/2018-12/Raytheon_article%20reprint_IDR%201901.pdf

[58] Office of the Director, Operational Test and Evaluation, "DOT&E FY 2016 Annual Report: Patriot Advanced Capability-3 (PAC-3)," December 2016, 175-177, https://www.dote.osd.mil/Portals/97/pub/reports/FY2016/army/2016patriot.pdf?ver=2019-08-22-105407-280 ; Isaac Maw, "Patriot Missile to Receive $133M in Upgrades Over Next Five Years," *engineering.com*, 9 July 2018, https://www.engineering.com/story/patriot-missile-to-receive-133m-in-upgrades-over-next-five-years

[59] Patrick O'Reilly, Ed Waters, "The Patriot PAC-3 Missile Program—An Affordable Integration Approach," https://apps.dtic.mil/sti/pdfs/ADA319957.pdf; Missile Defense Project, "Patriot," *Missile Threat*, Center for Strategic and International Studies, June 14, 2018, last modified March 24, 2022, https://missilethreat.csis.org/system/patriot/

[60] Missile Defense Advocacy Alliance (MDAA), "Patriot Advanced Capability-3 Missile, " 18 August 2020, https://missiledefenseadvocacy.org/defense-systems/patriot-advanced-capability-3-missile/ . This speed is consistent with a recent article giving a maximum speed of existing interceptors as "about 1.7 km/s" ("Japan set to develop railguns to counter hypersonic missiles," *NIKKEI Asia*, 4 January 2022, https://asia.nikkei.com/Politics/Japan-set-to-develop-railguns-to-counter-hypersonic-missiles ).

[61] North Atlantic Treaty Organization (NATO), "Patriot," Fact Sheet, December 2012, https://www.nato.int/nato_static/assets/pdf/pdf_2012_12/20121204_121204-factsheet-patriot-en.pdf ; MDAA, "Patriot."

[62] Leland H. Jorgensen, "Prediction of Static Aerodynamic Characteristics for Space-Shuttle-Like and Other Bodies at Angles of Attack from 0º to 180º," NASA Report TN D-6996 (1973), https://ntrs.nasa.gov/api/citations/19730006261/downloads/19730006261.pdf

[63] Jorgensen, "Prediction of Static Aerodynamic Characteristics."

[64] Leland H. Jorgensen, "A Method for Estimating Static Aerodynamic Characteristics for Slender Bodies of Circular and Noncircular Cross Sections," NASA Report TN 0-7228 (1973), https://ntrs.nasa.gov/api/citations/19730012271/downloads/19730012271.pdf

[65] Acton, "Hypersonic Boost-Glide Weapons;" "X-41 CAV (USAF/DARPA *Falcon* Program)," *Directory of U.S. Military Rockets and Missiles, Appendix 4: Undesignated Vehicles* (2009), http://www.designation-systems.net/dusrm/app4/x-41.html ; this is similar to the result found in Wright, "Research Note to Hypersonic Boost-Glide Weapons."

[66] Phillips, "A Common Aero Vehicle."

[67] Qinglin Niu, Zhichao Yuan, Biao Chen, and Shikui Dong, "Infrared Radiation Characteristics of a Hypersonic Vehicle Under Time-Varying Angles of Attack," *Chinese Journal of Aeronautics* 32 (2019): 867, https://doi.org/10.1016/j.cja.2019.01.003

[68] Using instead values for the CAV-L model gives somewhat lower lift but does not significantly change the results.

[69] Acton, "Hypersonic Boost-Glide Weapons."

[70] Candler and Leyva, "Computational Fluid Dynamics Analysis."

[71] U.S. Government Accountability Office, "Hypersonic Weapons: DOD Should Clarify Roles and Responsibilities to Ensure Coordination across Development Efforts," GAO-21-378
(22 March 2021): 13, https://www.gao.gov/products/gao-21-378 ; Steve Trimble, "Document Likely Shows SM-6 Hypersonic Speed, Anti-Surface Role," *Aviation Week*, 12 March 2020, https://aviationweek.com/defense-space/missile-defense-weapons/document-likely-shows-sm-6-hypersonic-speed-anti-surface-role ; Tyler Rogoway, "Navy To Supersize Its Ultra Versatile SM-6 Missile For Even Longer Range And Higher Speed," *The Drive*, 20 March 2019, https://www.thedrive.com/the-war-zone/27068/navy-to-supersize-its-ultra-versatile-sm-6-missile-for-even-longer-range-and-higher-speed

[72] Tracy and Wright, "Modelling the Performance."

[73] "Hypersonic and Ballistic Tracking Space Sensor (HBTSS)," *Missile Defense Advocacy Alliance* (2 July 2020), https://missiledefenseadvocacy.org/defense-systems/hypersonic-and-ballistic-tracking-space-sensor-hbtss/

[74] Missile Defense Agency (MDA), "MDA Hypersonic Concept," Defense Visual Information Distribution Service (6 June 2021), https://www.dvidshub.net/video/801628/mda-hypersonic-concept ; Theresa Hitchens, "Next Budget Will Limit Glide Phase Interceptor Contractors: MDA Head," *Breaking Defense,* 12 August 2021, https://breakingdefense.com/2021/08/next-budget-will-limit-glide-phase-interceptor-contractors-mda-head/

[75] The GPI is apparently intended to engage HGVs during their glide phase, which in principle would allow intercepts at longer ranges and could allow for defense of larger areas, relative to terminal phase defenses. Since it would engage BGVs before they were slowed by their terminal phase dives, the GPI would need to be considerably faster and more maneuverable than current endo-atmospheric interceptors. The desired design capabilities of GPI have not been publicly reported.

[76] See, eg, John D. Anderson, Jr., *Introduction to Flight, Eighth edition* (Reston, VA: American Institute of Aeronautics and Astronautics, 2016): 839.

[77] The glide altitude $h$ enters Equations A1 and A2 through $g$ and $V_e$ in the combination $R_e + h$, where $R_e$ is the radius of the Earth.

[78] Acton, "Hypersonic Boost-Glide Weapons."

[79] This result also comes from including the $V$ dependence of $\alpha$ and integrating the more exact Equations 30 and 31.

[80] Encyclopedia Britannica, "Stefan-Boltzmann Law," https://www.britannica.com/science/Stefan-Boltzmann-law

[81] Cameron L. Tracy and David Wright, "Modelling the Performance of Hypersonic Boost-Glide Missiles," *Science & Global Security*, vol. 28, no. 3 (2020): 135-170.

[82] Numerically integrating Equations (4) and (5) in our original equations simply adds up length increments at each time step in the local range and cross-range directions. We define angles $\Psi$ and $\Omega$ by dividing those lengths by $R_e$, but those angles do not appear in the Equations (1), (2), (3), or (6). They can be visualized as shown in Figure B1.

[83] "Spherical Trigonometry," Wikipedia, https://en.wikipedia.org/wiki/Spherical_trigonometry

[84] "Spherical Trigonometry," Wikipedia.